\documentclass{elsart}
\usepackage{epsfig}
\usepackage{epsf}
\def\etal{{\it et al}}
\newcommand{\CFO}{\mbox{C$_4$F$_8$O}}
\begin{document}
\begin{flushright}
SUHEP-10-2005\\
May 17, 2005\\
\end{flushright}

\begin{frontmatter}

\title{Performance of a C$_4$F$_8$O Gas Radiator Ring Imaging Cherenkov
Detector Using Multi-anode Photomultiplier Tubes }

\author{M. Artuso, C. Boulahouache, S. Blusk, J. Butt,}
\author{O. Dorjkhaidav, N. Menaa, R. Mountain, H. Muramatsu,}
\author{R. Nandakumar, K. Randrianarivony, R. Sia, T. Skwarnicki, S. Stone,}
\author{J. C. Wang and K. Zhang\thanksref{NSF}}
\thanks[NSF]{Supported by the National Science Foundation}
\address{Physics Department, 201 Physics Building}
\address{Syracuse University,  Syracuse, NY 13244-1130}

\begin{abstract}
We report on tests of a novel ring imaging Cherenkov (RICH)
detection system consisting of a 3 meter long gaseous C$_4$F$_8$O
radiator, a focusing mirror, and a photon detector array based on
Hamamatsu multi-anode photomultiplier tubes. This system was
developed to identify charged particles in the momentum range from
3-70 GeV/c for the BTeV experiment.

\end{abstract}

\end{frontmatter}

\section{INTRODUCTION}
Ring imaging Cherenkov (RICH) detection has proven to be a most
useful way to distinguish the identities of charged particles, both
hadrons and leptons, in high energy physics experiments
\cite{other-riches}. We report on measurements of Cherenkov ring
images using a 3 meter long octafluorotetrahydrofuran, C$_4$F$_8$O,
gas radiator and Hamamatsu multi-anode photomultiplier tubes
(MAPMT), model R8900-M16, to measure the Cherenkov angle. This gas
has never been used before in this application and the tubes are new
versions that also have not been tested. This system was designed
for the now defunct BTeV experiment planned for Fermilab. The main
design criteria was the ability to separate charged pions from kaons
at the four standard deviation level for momenta between 3 and 70
GeV/c. We intended to achieve this separation by detecting about 40
photons, each with angular resolution of 0.75 mr. We show how we
derive these expectations below.

The tests were done using a 120 GeV/c proton beam at Fermilab. We
constructed a full length version of the detector that had, however,
a smaller width and used a single spherical focusing mirror. The
full system is described in Ref. \cite{BTeV-RICH}.

This system was originally designed using C$_4$F$_{10}$ as a
radiator. The production of this gas, however, has been ended; the
last known supplier, the 3M corporation, has informed us that they
do not intend further manufacture. Thus although stockpiles of the
gas exist, the long term prospects for use in physics detectors are
not good. We will show that C$_4$F$_8$O provides an excellent
alternative.

\section{APPARATUS}
\subsection{Introduction}

\begin{figure}[htb]
\centerline{\epsfxsize 3.0in
\epsffile{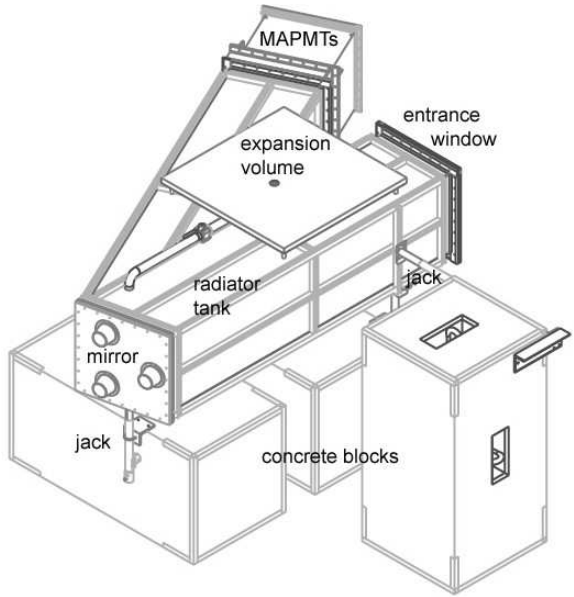}}
\caption{\label{setup} Schematic of the test apparatus. The beam
particles entered at the entrance window and traversed 3 meters of
gas. The Cherenkov photons were focussed by mirror onto the
MAPMTs.}
\end{figure}

The basic components of the test system are shown in
Fig.~\ref{setup}. The tank contains the gas in both arms. The beam
traverses 3 meters of gas in one arm in which it generates Cherenkov
light. The light is reflected by a spherical focusing mirror onto an
array of MAPMTs. These tubes are placed in their own enclosure so we
could work on them without opening the radiator gas volume. They are
separated from the gas volume by a 0.64 cm thick UVT acrylic window.
The MAPMT readout consists of specialized front end electronics
developed in conjunction with IDE-AS of Oslo, Norway. These devices
are mounted on hybrids connected a PCI based data acquisition system
described below.  We now discuss properties of each of these items.

\subsection{Mirror and Optics}
\begin{figure}[htb]
\centerline{\epsfxsize 3.0in \epsffile{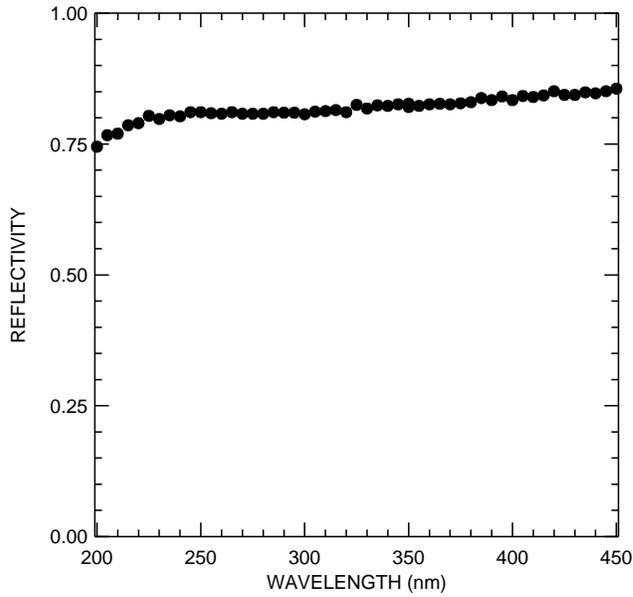}}
\vspace{-5mm} \caption{\label{mirror-ref} Measurement of mirror
reflectivity, in the center of the mirror, versus wavelength.}
\end{figure}

The mirror was produced by the COMPAS Consortium (Czech Republic).
It is made of glass with an aluminized front face having an
hexagonal shape 60 cm across each diagonal point-to-point. The
radius of curvature was measured to be 659 cm by finding the focal
position of the light from a small source reflected off the entire
mirror surface. The same technique allowed for measurement of the
spot size, defined as the diameter of the circle that contains 95\%
of the reflected light. For this mirror the spot size is 2.97 mm.
This is small enough that any dispersive effects due to mirror
quality will be negligible. The reflectivity of the mirror versus
wavelength is shown in Fig.~\ref{mirror-ref} \cite{thanks-brem}.
Note that the 81\% reflectivity around 300 nm is somewhat lower than
the 85\% specified by COMPAS and significantly lower than the 90\%
reflectivity measured for the HERA-b mirrors \cite{herab-mirror},
made using a similar technology.

\begin{figure}[htb]
\vspace{-0.8cm} \centerline{\epsfxsize 3.0in
\epsffile{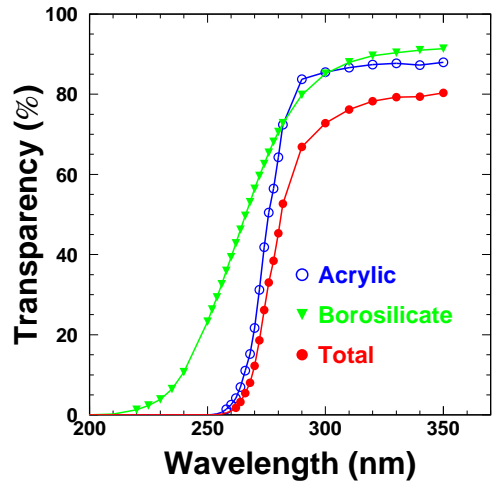}} \vspace{-0.8cm} \caption{\label{trans}The
transparency as a function of wavelength for 0.08 cm thick Hamamatsu
borosilcate glass (inverted triangles), and 0.37 cm thick Polymer
Plastics ultraviolet transmitting acrylic plate (open circles). Also
shown is the transmission through both for the actual used
thicknesses of 0.08 cm of glass and 0.64 cm of acrylic (filled
circles).}
\end{figure}

The transparency of a sample of the plastic used for the acrylic
window with a different thickness, 0.37 cm, is shown as a function
of wavelength in Fig.~\ref{trans} along with the transparency of a
0.08 cm thick sample of glass used in the MAPMTs. The glass
measurements are from Hamamatsu and repeated by us. Our measurements
were made using a spectrophotometer system, that is described in
detail in Appendix A.2 of ref. \cite{CLEO-RICH}.  The errors on the
measurement are $\pm$1\%, absolute, at all wavelengths. Our system
is designed to work in the wavelength region from approximately 280
nm to about 600 nm, where the quantum efficiency of the MAPMTs falls
to zero. The product of the acrylic window and MAPMT window
transmission is also shown on Fig.~\ref{trans}, where the
transmission for the actual material thickness are used.

\subsection{MultiAnode-PhotoMulitplierTubes}

\begin{figure}[htb]
\center{\epsfig{figure=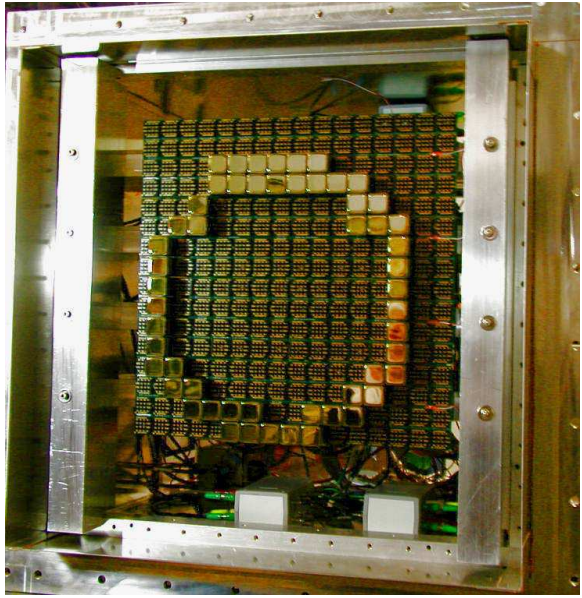,
width=3in}} \caption{\label{mapmt} A photograph of the MAMPT
system mounted on the high voltage baseboards.}
\end{figure}

Our 53 MAPMTs were arranged in an approximate circle, as shown in
Fig.~\ref{mapmt}, in order to intercept most of the Cherenkov ring.
A few were purposely displaced to be able to monitor noise or
backgrounds. The tubes and the electronics were water cooled. The
temperature varied from 26$^{\circ}$-33$^{\circ}$C, depending on the
vertical position and was kept constant within $\pm$0.2$^{\circ}$C
when running at our nominal voltage settings.

\begin{figure}[htb]
\center{\epsfig{figure=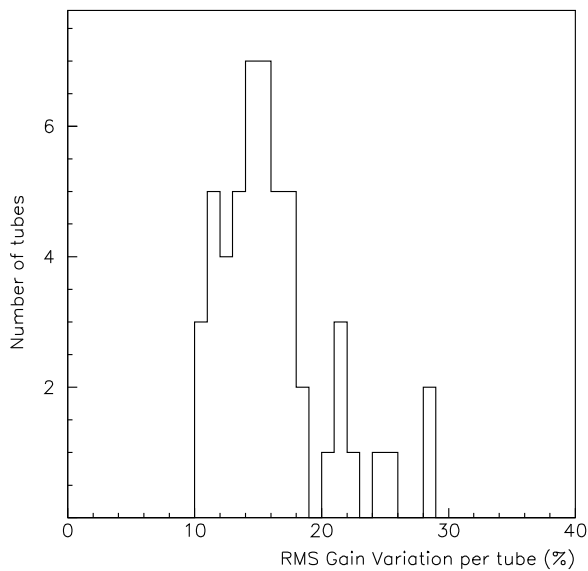, width=3in}} \vspace{2mm}
\caption{\label{rms_gain} The r.m.s. gain variations within each
tube.}
\end{figure}

\begin{figure}[hbt]
\centerline{
\includegraphics[width=3.2in]{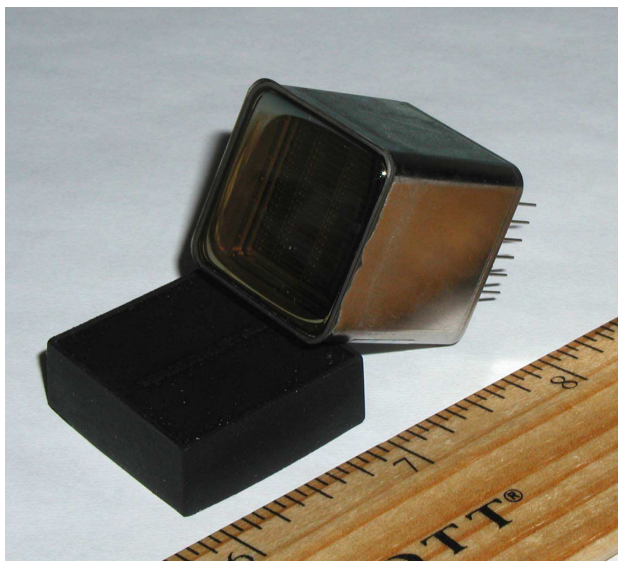}}
\caption{\label{fig:mapmt_photo} Photograph of a R8900-M16 MAPMT.}
\end{figure}

\begin{figure}[hbt]
\vspace{-1in} \centerline{
\includegraphics[angle=-90,width=6.0in]{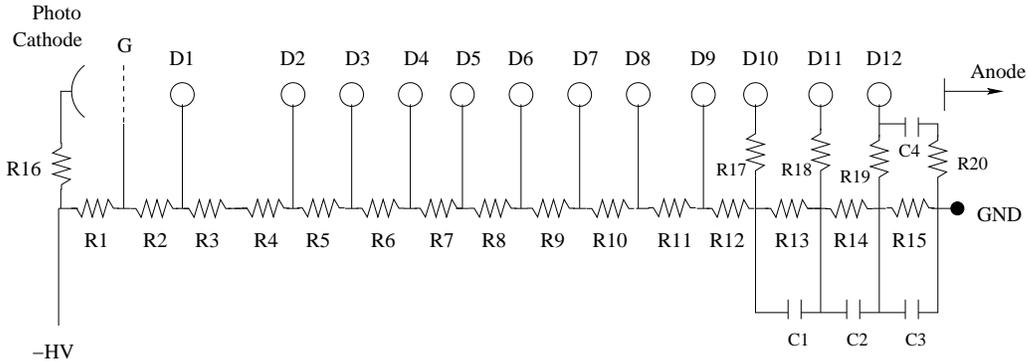}}
\vspace{-1.3in} \caption{\label{fig:vdn} Electrical schematic of the
voltage divider network. Resistors (R) and capacitors (C) are
indicated. The D's indicate dynodes, and G is the focusing grid.}
\end{figure}

    The R8900-M16 multianode photomultiplier tube (MAPMT) is a
26.2 mm $\times$ 26.2 mm square PMT with a 0.8 mm thick borosilicate
glass window and a Bialkali (K-Cs) photocathode. The active area of
the photocathode is specified to be 24 mm $\times$ 24 mm. The tube
is segmented into a 4 $\times$ 4 array of independent photodetection
elements. Each cell has its own 12-stage dynode chain connected to
an anode output. A photograph of the MAPMT is shown in
Fig.~\ref{fig:mapmt_photo}. A single HV line and a voltage divider
network (VDN) provide voltages to the photocathode and the dynodes.
A schematic of the VDN is shown in Fig.~\ref{fig:vdn}. In our
initial studies of the tubes, and for the results presented in this
section we used resistances of 110 k$\Omega$, 330 k$\Omega$, 330
k$\Omega$, and 110k$\Omega$ for R1-R4, 180 k$\Omega$ for R5-R15, 1
M$\Omega$ for R16, 51 $\Omega$ for R17-R19, and 20k$\Omega$ for R20,
as recommended by Hamamatsu. The capacitors, C1-C4 are all 0.01
$\mu$F, 200 V. Gains for individual tubes varied in the range
1-4$\times~10^6$ at 800 V. More important for our purposes is the
gain variations within a tube, since they can't be reduced by tuning
voltage settings. Fig.~\ref{rms_gain} shows the r.m.s. variations
within each of our tubes. The channel-to-channel gain variation
within a tube, averaged over all tubes, has an r.m.s. width of 16\%.

For the test beam run we decided to run at lower voltages in order
to have less gain. In order to maintain full collection efficiency,
we use a different set of resistances in the first four stages
(R1-R4) of 180 k$\Omega$, 536 k$\Omega$, 536 k$\Omega$, and
180k$\Omega$, the rest staying the same. These voltage dividers were
implemented on baseboards that host 16 MAPMT's each. The baseboards
also transmit the MAPMT signals via flat multiconductor cables to
hybrid boards that host the electronics.

    The R8900 is the result of an evolution of MAPMTs from Hamamatsu which
began with tubes with active area as low as 36\% (R5900) and
culminated with this tube which is designed to have an active area
approaching 85\%.

\begin{figure}[hbt]
\centerline{
\includegraphics[width=6.0in]{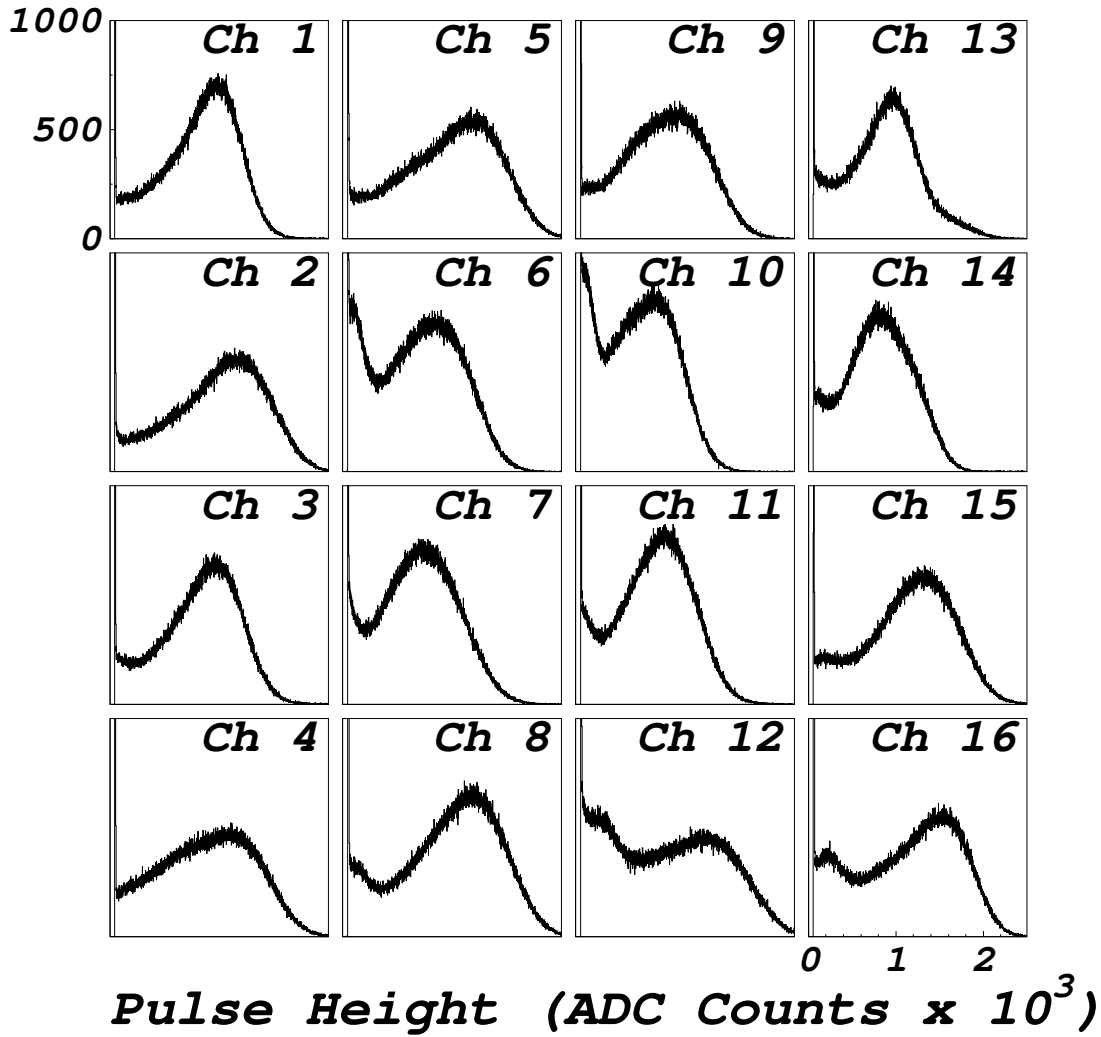}}
\caption{\label{fig:ph_spectrum} Pulse height spectra for one of the
R8900 MAPMTs, where all of the 16 channels are shown in their proper
geometrical positions. (All the horizontal and vertical scales are
the same.)}
\end{figure}

We performed a series of bench tests including: comparison of the
pulse height spectra from channel-to-channel, determination of the
active area, and measurement of the sensitivity to external magnetic
fields. These tests were performed using a low-intensity blue LED,
wavelength of 470 nm with a spread having a full width at half
maximum of 40 nm, whose light was transported inside a dark-box
along a single-mode optical fiber. The tip of the fiber was pointed
at the center of each channel and the pulse height spectra was
measured using a PCI-based ADC card. The voltage was set to 850 V
for these initial measurements. The pulse height spectra for one of
the MAPMTs are shown in Fig.~\ref{fig:ph_spectrum}. The
channel-to-channel gain variations are at the level of 15-20\%. Most
of the channels have a well-defined peak and valley, although
channels 6, 10, and 12 have a larger fraction of low pulse height
signals. This is not a significant problem since the VA\_MAPMT chip
(see Section~\ref{sec:elect}) which is used for the MAPMT readout
can operate at low threshold.

    To investigate the active area of the tube and the magnetic field sensitivity,
the signals from the MAPMT were read out using CAMAC and a
LabVIEW-based DAQ. The CAMAC system includes an amplifier (LeCroy
\#612A), discriminator (Phillips Scientific, \#710) and scaler
(LeCroy \#2551). The threshold was set to a relatively high value,
which resulted in some clipping of low pulse height signals,
although for these studies, we were mainly interested in relative
response.

\begin{figure}
\centerline{
\includegraphics[width=5.5in]{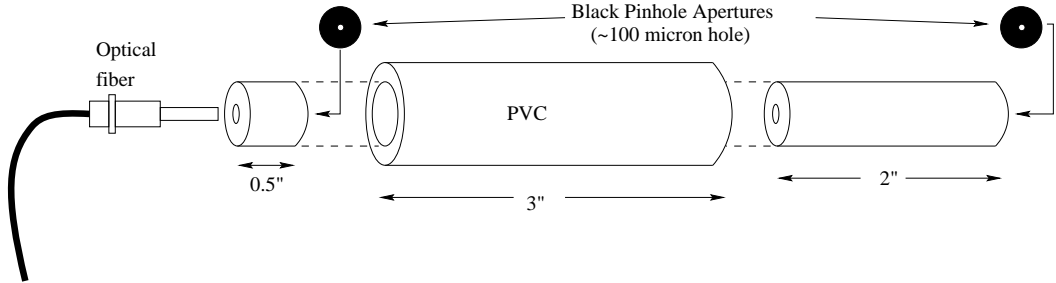}}\vspace{2mm}
\caption{\label{collimator} A sketch of the collimator used for the
MAPMT scan.}
\end{figure}

\begin{figure}[hbt]
\centerline{
\includegraphics[width=5.5in]{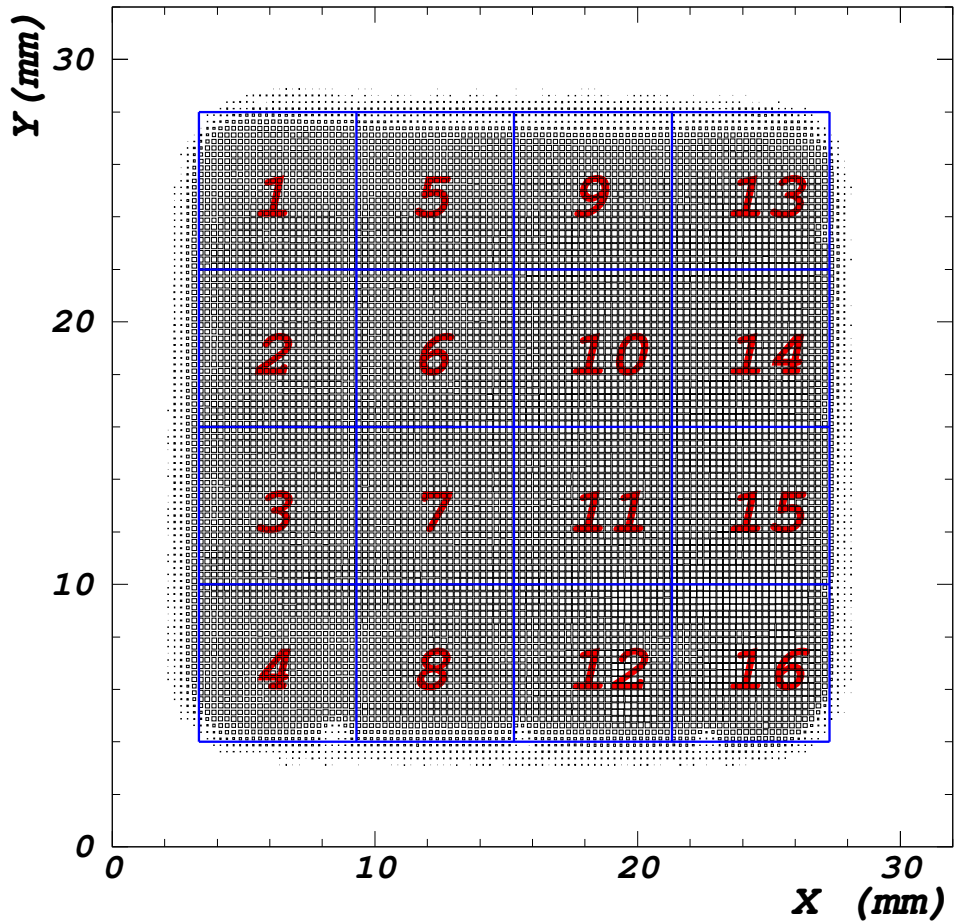}}
\caption{\label{fig:xyscan_na0061} Two dimensional scan of an
R8900-M16 MAPMT. Both the X and Y step sizes are 0.25 mm. The
channel geometry is shown in the boxes. The data have not been
background subtracted (see text).}
\end{figure}

    For the active area determination, we scan over the face of the MAPMT
in steps of 0.25 mm in both X and Y. The light from the fiber was
passed through an additional collimator which provided, on the face
of the tube, an approximately Gaussian shaped spot in both X and Y
dimensions with an r.m.s. width of about 150 $\mu$m. A sketch of the
collimator is shown in Fig.~\ref{collimator}. The tubes are
positioned 1 mm away from the tip of the fiber with an accuracy of
0.5 mm. The fiber position is determined to 20 $\mu$m precision in
both X and Y and the rotational accuracy is 20 mr. At each point we
measure the total number of counts above the discriminator threshold
in a time interval of 2 seconds. The dark count rate was typically
about 40 Hz and was therefore neglected. The results of a typical
scan are shown in Fig.~\ref{fig:xyscan_na0061} where we plot the
integrated count rate as a function of the (X,Y) coordinates of the
spot. The outline of the tube is evident. A 24 mm square box is
superimposed, which shows that the active area in X is about 24 mm,
while in Y it is closer to 23 mm. The asymmetry of the dynode
structure may account for this difference. These data indicate that
80\% of the tube's physical area is active. Counts outside the box
are due to background which has not been subtracted.

\begin{figure}[hbt]
\centerline{
\includegraphics[width=6.0in]{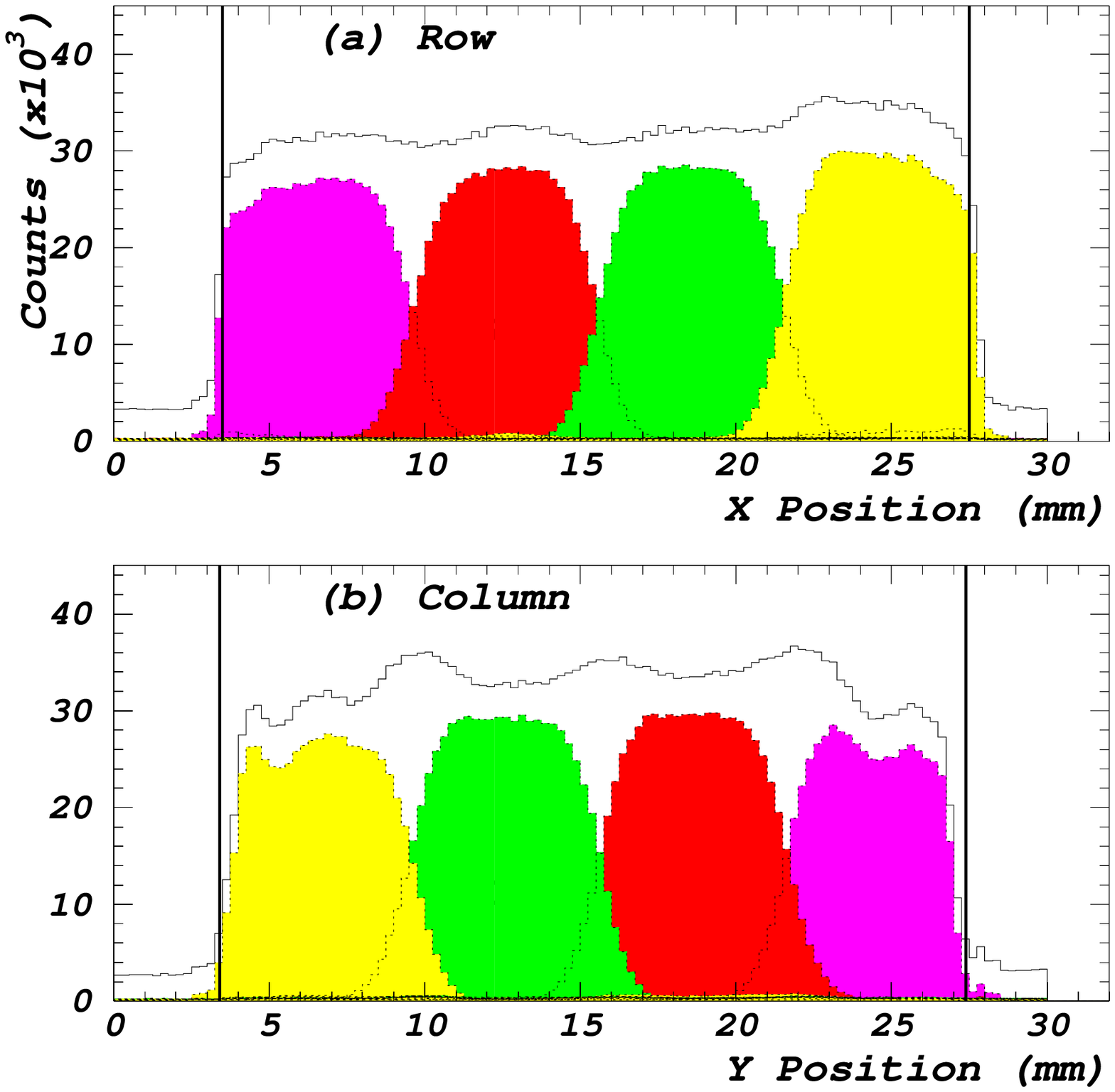}}
\vspace{-0.3in} \caption{\label{fig:rowcol_na0061} Slice, 0.25 mm
wide, of the two-dimensional XY scan along 4 rows of the R8900
passing through the center of the channels. Step sizes are 0.25 mm
in X and Y. The unshaded histogram is the sum of all channels while
the shaded histograms correspond to individual channels.}
\end{figure}

The individual channel responses are more clearly illustrated in
Fig.~\ref{fig:rowcol_na0061}, where we project out a slice of
Fig.~\ref{fig:xyscan_na0061} along (a) X and (b) Y, such that the
slice passes through the center of a row and a column.  The unshaded
histogram shows the integrated count rate for all channels and the
shaded histograms show the contributions from the individual
channels in that particular row or column. The sharp drop off at the
edge of the photocathode confirms the small spot size. Near the
border of neighboring channels, we observe a region of ambiguity
where a photon can be focused onto the first dynode of either
channel. Once in that channel, it is expected that the signal will
be collected on the corresponding anode. To quantify this effect, we
fit the edge to an error function and extract an r.m.s. width of
0.75 mm, after a small correction for the size of the light spot
($\sim$0.15 mm). Alternately, we compute the fraction of counts
which falls outside the geometric edge of a channel relative to the
total counts within the 6 mm x 6 mm boundary. We find that about 5\%
of the counts spill over into a neighboring channel for each edge.
Therefore, for an interior channel, about 20\% of the total counts
are detected on neighboring channels; this corresponds to 5\% for
each of the four neighbors. These photons are generally close to the
periphery, as indicated by the roll off at the edge of the channels
(see Fig.~\ref{fig:rowcol_na0061}). This leads to a small
degradation in the positional resolution of this device relative to
one which has no spill over.

Lastly, we report on measurements of the magnetic field
sensitivity of the R8900-M16 MAPMT. For use in BTeV, we required
the tubes to perform well in an external field as large as 10
Gauss. We studied two corner channels (channels 1 and 13), one
non-corner edge channel (channel 3) and one interior channel
(channel 10). (For some tests we included another corner channel
16.) The tests consisted of measuring the count rate from a pulsed
LED with zero field and then with a known field provided by a pair
of Helmholtz coils. We investigated both the orientation parallel
to the axis of the tube (longitudinal field)and perpendicular
(transverse field).

\begin{figure}[hbt]
\centerline{
\includegraphics[width=5.0in]{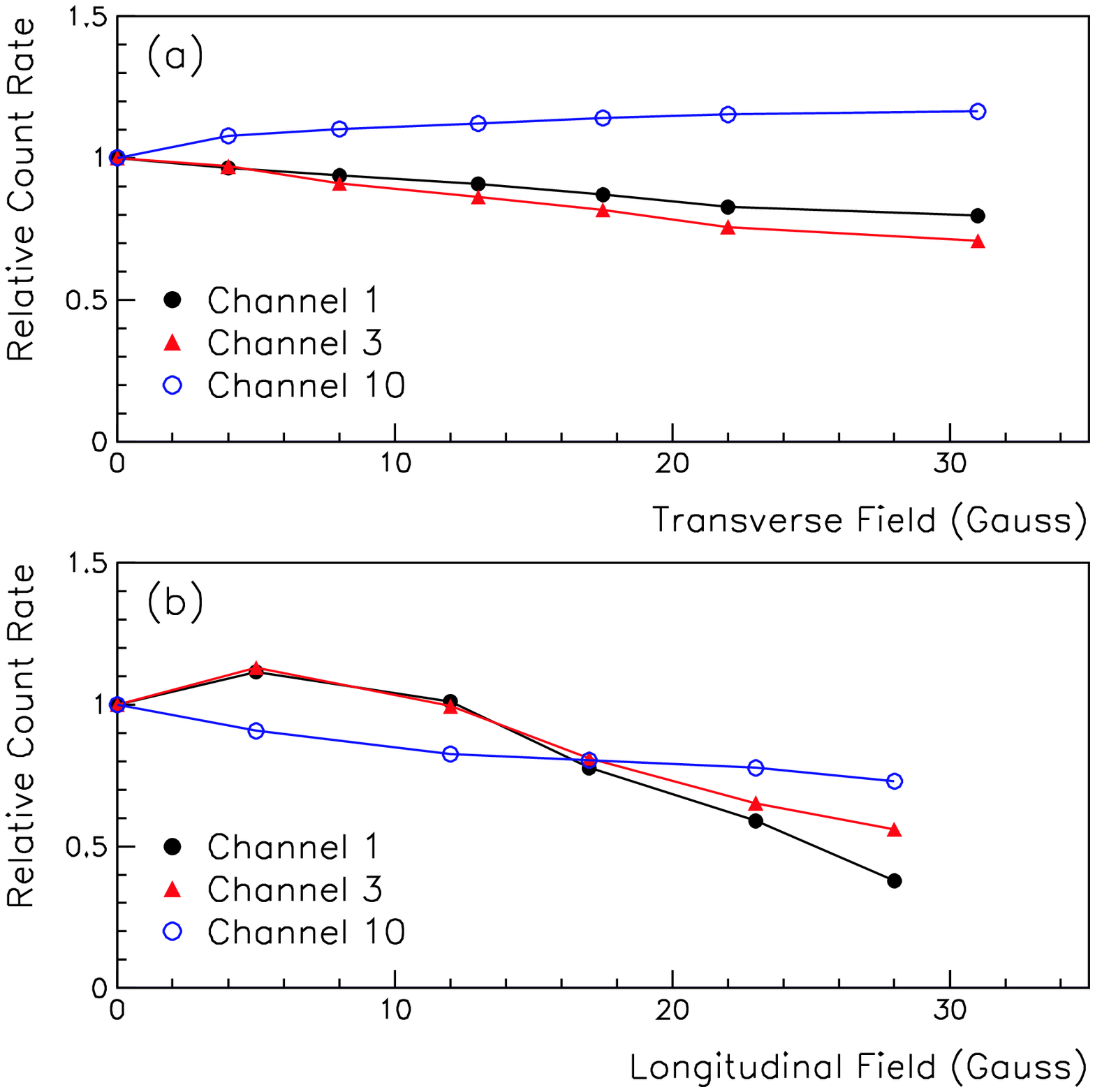}}
\vspace{.2in} \caption{\label{mapmt_noshield} Count rate as a
function of the applied magnetic field without external shielding
for (a)) transverse field and (b) longitudinal field.}
\end{figure}

The results of applying a longitudinal magnetic field on the tube
without shielding are shown in Fig.~\ref{mapmt_noshield}. The
efficiency was mildly dependent upon the transverse field, dropping
by $\sim$10\% with a transverse field of 15 Gauss for a corner
channel, and by less than 5\% for interior channels. On the other
hand, significant losses in collection were observed when
longitudinal fields were applied.

\begin{figure}[hbt]
\centerline{
\includegraphics[width=6.0in]{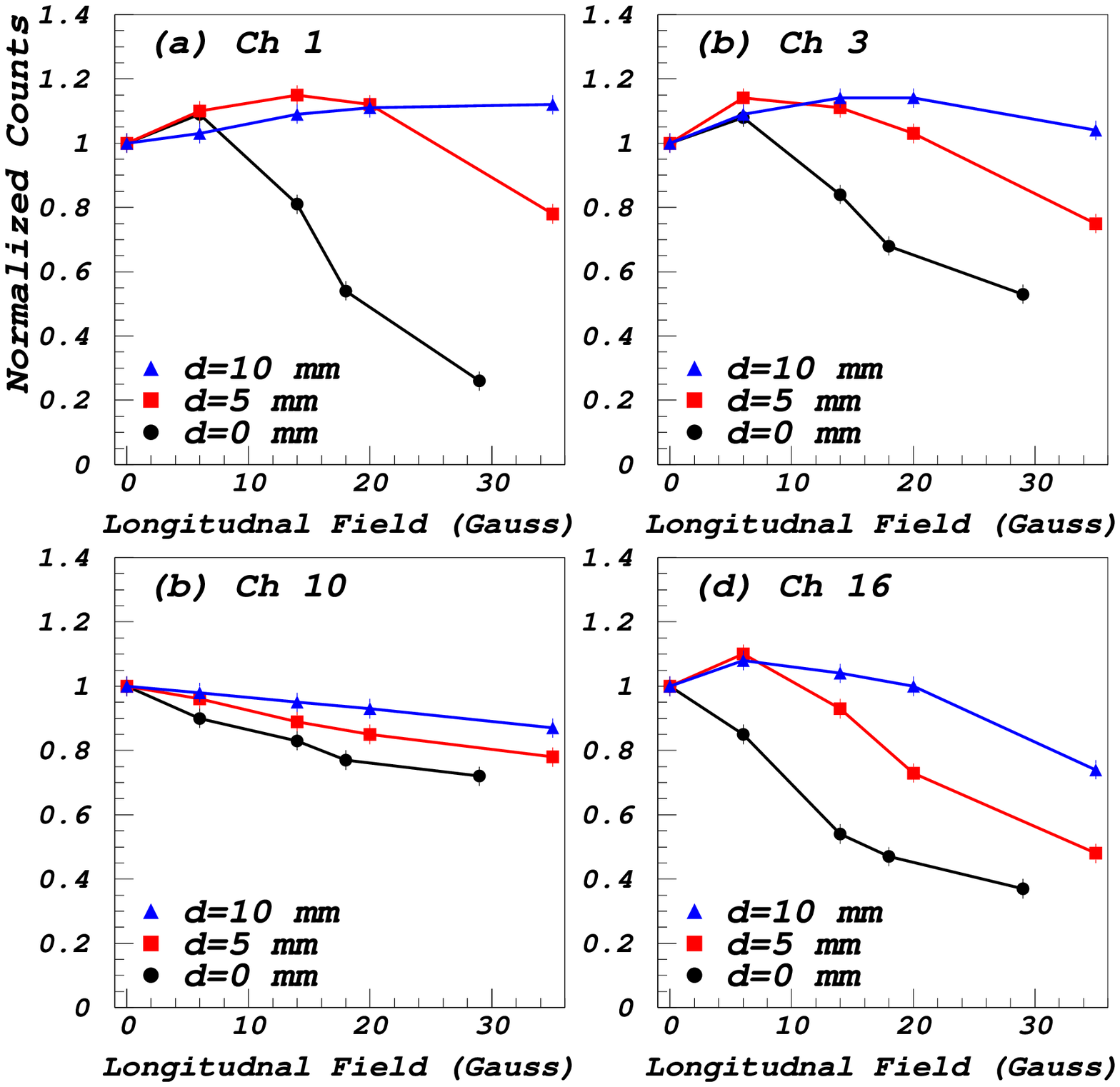}}
\vspace{-0.2in} \caption{\label{fig:mag_sens} Count rate as a
function of the applied longitudinal magnetic field for shield
extension values, $d$=0 mm, 5 mm, and 10 mm, for (a) Channel 1, (b)
Channel 3, (c) Channel 10, and (d) Channel 16.}
\end{figure}

We investigated shielding the tube with a 1 inch long 250 $\mu$m
thick mu-metal shield. The shield was electrically isolated from the
outer can of the MAPMT and was allowed to float electrically. Count
rate measurements were taken with different extension lengths, $d$,
of the mu-metal tube beyond the face of the MAPMT. Data were taken
for $d$=0 mm, 5 mm, and 10 mm. The results are shown in
Fig.~\ref{fig:mag_sens} for channels 1, 3, 10, and 16. Each set of
data are normalized to the count rate at zero applied field. We find
that when the shields are not extended ($d$=0 mm), the corner and
edge channels incur large losses, coming from both collection
efficiency and a decrease in gain; in fact, there is almost no
shielding effect in this configuration. As the shielding is extended
the losses are reduced, and at 10 mm, the drop in efficiency (at 10
Gauss) is typically less than 10\%. Some channels show an increase
in count rate as compared to no field, and hence the average loss is
typically less than 5\%.

\subsection{Beam Properties}

The beam test data was taken at Fermi National Accelerator
Laboratory, in the Meson Test Beam Facility (MTBF) \cite{Ram04}. The
RICH detector was set up in the MT6B enclosure. The beam as
delivered was 120~GeV protons, from the resonant extraction of a
single Booster batch in the Main Injector, with a $<$4\%\ momentum
bite. The beam tune was varied somewhat during the several days of
data-taking, but typical beam rates were 5 KHz per spill, as
measured by beam counters.
Spills lasted 700~ms, with 1 to 8 spills per supercycle (60~s). The
data-taking runs lasted on the order of 30~mins.

\subsection{Electronics, Trigger and Data Acquisition}
\label{sec:elect} The MAPMTs are plugged into a custom PCB baseboard
that contains the VDN's and signal routing. Each baseboard hosts 16
MAPMTs, a total of 256 channels, and passes the signals along a
34-conductor cable to two hybrids one above and one below the
support channel. Each hybrid hosts two 64-channel ASICs, thus
accommodating 8 MAPMTs.

The front end electronics used in the test beam is based on custom
made VA\_MAPMT ASICs produced at IDE AS, Norway.{\footnote{Ideas
ASA, N-1330 Fornebu, Norway; http://www.ideas.no.}  Multiplexer
boards (MUX), designed and developed at Syracuse, provide the
interface with a PCI based data acquisition system designed and
produced at Fermilab \cite{lorenzo}. A system diagram describing one
of the readout chains is shown in Fig.~\ref{tb_cartoon}.

\begin{figure}[hbt]
\center{\vspace{-0.9in} \epsfig{figure =
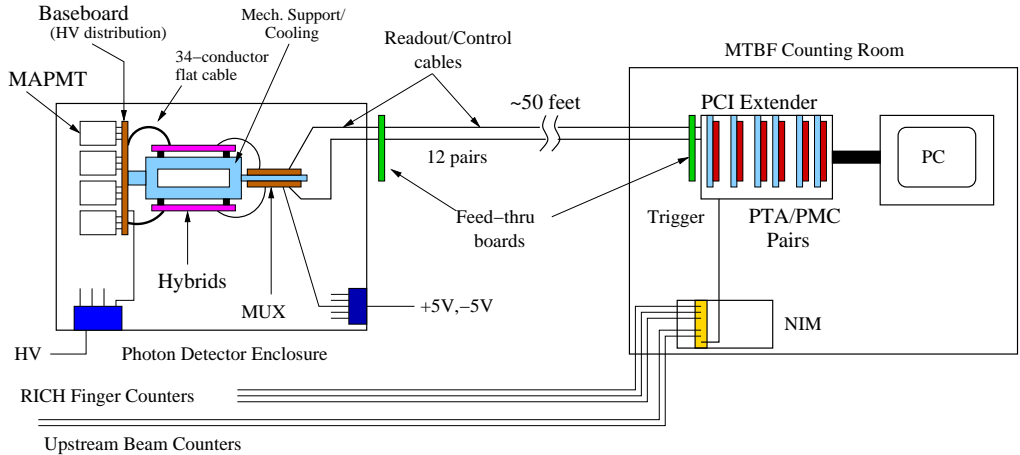,angle=-90,width=5.5in}
}\vspace{-0.9in} \caption{Diagram showing major components of
electronic read-out system.} \label{tb_cartoon}
\end{figure}

The key elements in this data acquisition system are contained on
pairs of mating boards. The mother-board (PTA) is connected to a
LINUX PC via a PCI bus. It contains the basic functionality to
interface the online processor to the front end system, a sizeable
memory to store the events and a variety of registers that allow the
customization of the readout procedure to a variety of different
systems. The mating board (PMC) contains a large FPGA (Xilinx Virtex
PRO) that permits customization of the DAQ. We used the general PTA
firmware developed for the BTeV pixel test beam \cite{lorenzo}, but
wrote the PMC firmware suitable for our data architecture.

Two finger size scintillation counters were placed in front of the
RICH tank in the horizontal and vertical directions and an
additional one arranged vertically was placed behind the tank. They
were aligned to beam particles. We also used two larger beam
defining counters upstream.  Our trigger was formed as a coincidence
of the beam counters and local finger counters. The trigger signal
required the coincidence of all five scintillators. The typical
trigger rate was 300-500 Hz, during data-taking. The beam spot at
the entrance window of the RICH prototype tank was $\sim$5~mm in
diameter. The r.m.s. angular divergence of the beam was
$<$0.25~mrad. This trigger signal was used to move the data from the
local buffers in the front end hybrids to the PTA memory.

\subsubsection{Front End Electronics}
The VA\_MAPMT ASICs comprise 64 mixed analog and digital processors
with parallel inputs and parallel outputs. The parameters affecting
its mode of operation are loaded through an input shift register
initialized at the beginning of a data taking cycle.

Fig. \ref{fig:blockdia}  shows the conceptual diagram of each
readout channel. The analog section comprises a semi-gaussian
preamplifier and shaper circuit, followed by a high pass filter that
has the purpose of reducing the sensitivity of the discriminator to
long range drifts of the DC working point of the device. In
addition, a voltage-controlled pole-zero cancellation circuit is
introduced to optimize the rate capabilities. The analog section is
optimized for low noise performance with an expected gain of $10^6$.

\begin{figure}[hbt]
\center{\vspace{-0.0005in}
\epsfig{figure=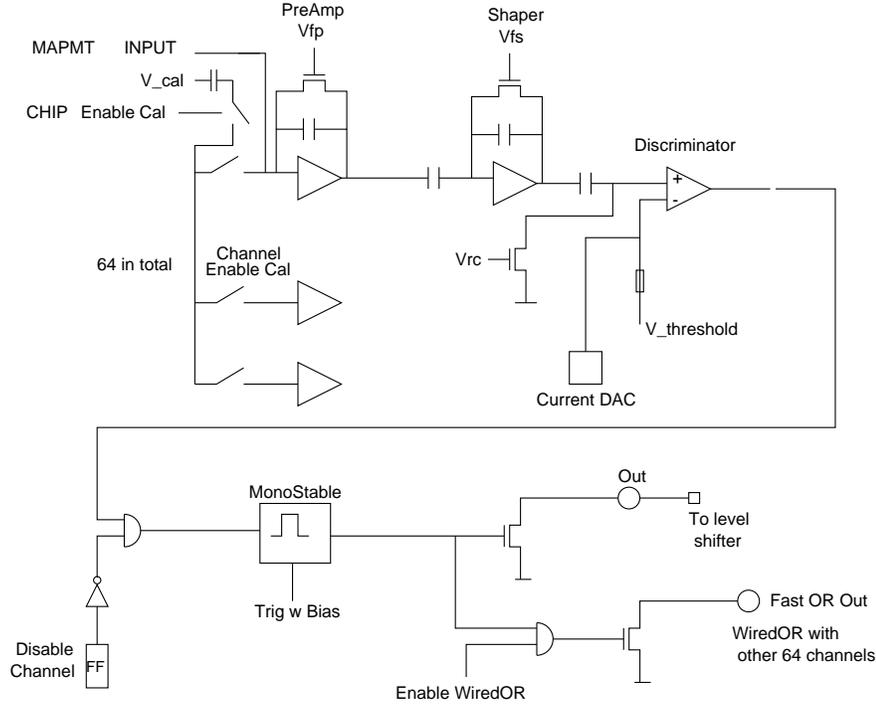,width=5in} }\vspace{-0.0005in}
\caption{ Block diagram of an individual readout channel of the
front end ASICs described in this paper.} \label{fig:blockdia}
\end{figure}

The input of the digital section is a discriminator that must
operate effectively at very low thresholds and it needs to tolerate
very high rates, of the order of several MHz, to cope with the high
occupancy that was expected in some areas of the BTeV RICH. The
discriminator threshold is set through an external 8 bit DAC. In
addition, a 4 bit programmable DAC is built in every channel to fine
tune the threshold of each individual channel to compensate for
different DC offsets. The discriminator output drives a monostable
circuit that produces an output current pulse whose width is about
100 ns. Individual digital outputs can be disabled through a channel
mask set during the initialization sequence.

There are three modes of operation for this ASIC: (1) an
initialization sequence, when a bit pattern sequence is shifted in
the ASIC to program the desired operating conditions; (2) a
calibration mode, when channels selected in the initialization
sequence respond to an input current pulse sent to the calibration
input; (3) finally, in normal mode, all the working channels are
activated and respond to charge signals collected at their inputs.
In addition, a fast-OR of all the channel hits can be activated for
monitoring or synchronization purposes.

Fig.~\ref{mapmt-hyb} shows the hybrid hosting the VA\_MAPMT ASICs.
It is a conventional 6-layer rigid printed circuit board. The 64
parallel current outputs of the VA\_MAPMT ASIC are wire-bonded to
the inputs of a level-shifter ASIC that produces TTL logical signals
matching the input requirements of the XILINX Virtex 300 FPGA, used
to drive the initialization sequence and they latch and transfer the
data from the front end to the back end circuit with the protocols
needed by the data acquisition system. The firmware is downloaded in
the first step of the initialization sequence and thus we can adapt
this hybrid to different data taking modes, and different triggering
configurations.

\begin{figure}
\center{ \vspace{-2.5in}\epsfig{figure
=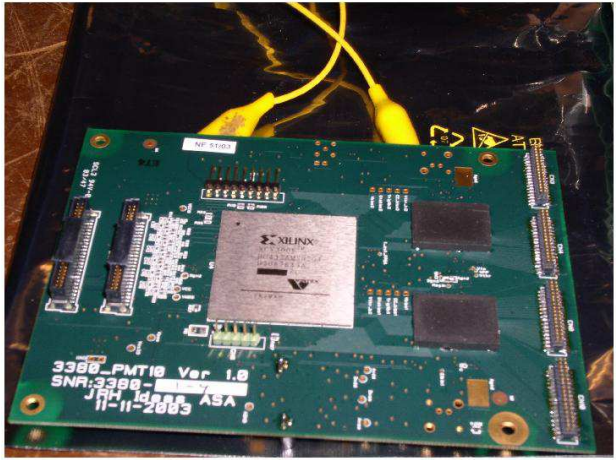,width=3.5 in}
} \caption{ VA\_MAPMT ASICs mounted on the hybrids used in the
BTEV gas RICH prototype studied with test beam runs at Fermilab.}
\label{mapmt-hyb}
\end{figure}

\subsubsection{Data acquisition architecture}

There are two firmware modules that were developed for this test
beam to allow for triggered data acquisition: a hybrid firmware,
implementing the ASIC initialization and the timing necessary to
synchronize the data flow with the trigger signal, and a PMC
firmware, to interface this data structure with the general data
acquisition framework.

Two elements needed to be implemented in the hybrid firmware: a wait
period to compensate for the delay between the scintillator trigger
and the prompt binary information appearing a few ns after a charge
signal exceeded the given threshold. The binary output was only 100
ns wide, to allow for the fast data rates expected in the real
experiment, and thus it decayed before the corresponding trigger was
available at the hybrid boards. Secondly, a time stamp mechanism was
necessary to allow the combination of event fragments from different
hybrids.

Channels registering a signal above threshold self-latched and
initiated a delay signal matching the expected trigger delay. At the
end of this wait period, if a trigger signal was asserted, the data
were pushed out of the hybrid onto the PMC board, whereas if no
trigger occurred, the latches were reset and the hybrids were ready
for a new event. When a channel was latched, the corresponding 6-bit
event counter information was latched in the hybrid. The event
counters in the various hybrids were driven by a global
synchronization clock provided by one of the PMC cards, the master
card. The master card distributed the clock to the other cards
which, in turn, distributed it synchronously to all the front end
hybrids.

The PMC cards received the signals coming from 2 hybrids. Their main
functions were the generation of an ``extended time stamp," using a
26 bit counter synchronized with the local time stamp counter.
Subsequently the formatted events, including the hybrid address,
were sent to the online PC for further processing.

The DAQ software \cite{magni}, originally developed for the BTeV
pixel beam test, was implemented in C++ and encompasses three main
processes: Producer, Consumer and Logger. The Producer is the
component which works mostly with the PTA/PMC pair. It sends the
control signals necessary to drive the data acquisition and receives
data from the local memories. The Consumer process reads and
analyzes the data. The Logger receives messages from the other
components and notifies the user in case of errors.


\subsection{Gas System}

A gas system was designed and built to circulate and purify the
\CFO\ radiator gas. The design was intended to be open and passive,
not requiring complicated feedback with precise pressure monitoring.
As a consequence, an expansion bellows was mounted above the main
RICH tank, which expanded to take up the volume of a gas charge when
injected into the system, and compensated for variation in
atmospheric conditions. A simplified schematic of the system is
shown in Fig.~\ref{fig:gasschematic}.

There are two main circuits in the system. The first is the
compression circuit, which consists of the main RICH tank, expansion
bellows, high-pressure pump, small condensation tank, \CFO\ gas
bottle, and a filter pack. This is the outer circuit shown in the
flow diagram. Initially the system is flushed with Argon. Filling
with \CFO\ gas is accomplished by pressure differential flow from
the bottle through the filters to the main tank.  The flow rate is
controlled so as not to stress the acrylic windows on the main tank,
and the expansion bellows expands to hold the extra gas volume. When
the bellows reaches its maximum extent, the large pump is turned on,
and the gas (a mixture of Argon and \CFO) is compressed in the small
condensation tank.  When the pressure in the condensation tank
reaches $\sim$25~psig, the \CFO\ gas liquifies; and when the
pressure reaches 33~psig, a relief valve opens releasing the gas
(mostly Argon) until the pressure falls below the relief level. In
this way, the fluid mixture is purified of Argon (as well as any
residual air).

A new \CFO\ gas charge is injected from the bottle, passing through
the filter pack, and into the main tank, again filling the expansion
bellows. The filter pack consists of a coalescer, a molecular sieve
and a particulate filter to remove water vapor and particulates from
the gas. This compression process is repeated iteratively until the
required purity is reached. After a few cycles, a valve is opened to
allow the \CFO\ liquid in the compression tank to vaporize and
re-enter the circuit as a new charge. In this beam test,
$(94\pm2)$\% purity by weight was achieved asymptotically. This was
the maximum possible in the system, limited by the existence of a
small volume at the top of the condensation tank that was
inaccessible to liquid. The remaining gas in the system was argon
that is transparent in the wavelength range of interest.

The second circuit in the system is the circulation circuit, which consists of the main tank,
expansion bellows, filters, and a small pump.  It is the inner circuit shown in the flow diagram.
This circuit is run when the tank is filled with sufficiently
purified \CFO\ gas, and is intended to maintain the radiator gas at a constant composition
during data taking.

The circuits were controlled and monitored by the PLC-based iFix
DMACS$^{\rm TM}$ system from General Electric-Intellution, which
provided the computer interface for controlling the pumps and
reading the transducers for pressure, flow, temperature and
humidity. A height sensor on the expansion bellows allowed for a
simple feedback loop to control the high-pressure pump.  All process
variables were recorded.
Overall the \CFO\ gas system worked well, as designed. We did find,
however, that the expansion volume while useful for a test system
would be too cumbersome to include in the final detector. Rather the
actively controlled gas systems similar to those used on the DELPHI
\cite{DELPHI} or HERAb \cite{herab-mirror} RICH systems would be
more appropriate.

\begin{figure}[tb]
  \vspace{0.1in}
  \begin{center}
    \includegraphics*[height=5.2in]{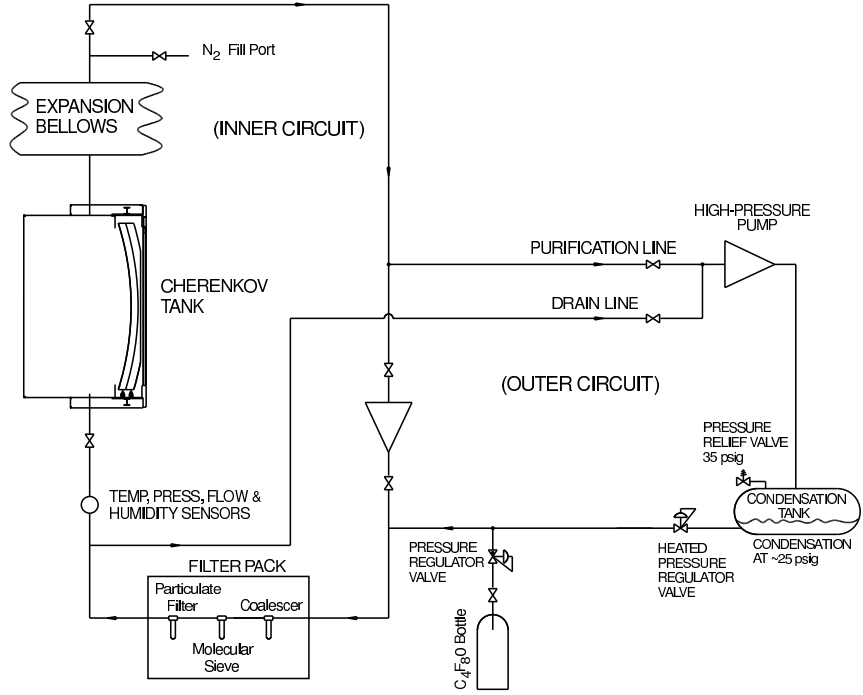}
  \end{center}
  \vspace{-1.5in} \caption{\label{fig:gasschematic} Simplified flow
  diagram of the \CFO\ radiator gas
    circulation and purification system.  }
  \vspace*{-0.001in}
\end{figure}


\subsection{\CFO\ Gas Compatability}

Since \CFO\ has never before been used as a Cherenkov radiator, its
chemical compatability with other materials needed to be
investigated. To this end, \CFO\ gas was tested in contact with a
variety of plastics, epoxies, metals and other materials, as listed
in Table~1 \cite{Nand04}.

The \CFO\ gas and all materials (except water) were put into the
same specially-designed gas-tight stainless steel vessels and kept
at an elevated temperature (80$^\circ$C) and pressure (1.2~atm) for
several months to accelerate the chemical kinetics of any possible
reactions. This represented a worst-case test.
After periods lasting up to 9.6 years exposure (20$^\circ$C
equivalent), the \CFO\ gas was analyzed via Gas Chromatography and
Proton NMR. No measurable amounts of reaction products were
obeserved. The materials exposed were tested for relevant physical
properties: change in dimensions, weight, epoxy strength, and
optical transmission (e.g., UVT acrylic). No measurable material
changes were detected,\footnote{
    There was a slight effect, however, in lightening the color of the surface
    of the Copper tube after 4 years equivalent exposure.
    Electron-dispersive X-ray analysis did not indicate the presence of
    Fluorine on the Copper surface to a depth of 1~$\mu$m.
    A control vessel filled with dry N$_2$ exhibited a similar result.
    This is most simply understood as a temperature induced effect.}
and no optical changes were seen larger than the sample-to-sample
variation ($<$2\%\ above 290~nm, $<$5\%\ below).

Water is a special concern for \CFO, since the Oxygen atom was
thought to make it potentially reactive, forming HF. However, in a
separate test with only water exposure to \CFO\ gas for up to 13.5
years equivalent, gas chromatography analysis found the total
contaminants picked up by the gas to be $<$50~ppm (resolution of the
device), and proton NMR studies gave no indication of any H-based
contaminants. There was no measurable HF production.

\begin{table}[t]
\centering \label{tab:materials} \caption{ Materials tested for
compatability with \CFO. }
\begin{center}
\begin{tabular}{ll}
 Class & Materials \\
\hline\hline
 Possible reactants & Water \\
 Window materials & UVT Acrylic, Borosilicate Glass \\
 Mirror materials & Continuous Fiber Reinforced Plastic (CFRP), Mirror matrix \\
                  & Beryllium metal, Beryllium with CYTEC BR-127 primer \\
 Construction metals & Aluminium, mild Steel, Stainless Steel, Copper, Brass \\
 Composite materials & Teflon, Kevlar, Carbon Fiber, FR-4 \\
 Epoxies             & Armstrong A-12, Hysol, Araldite \\
 O-Ring materials    & Viton, Buna-N, Neoprene \\
 Other materials  & Poly-flo tubing \\
\hline
\end{tabular}
\end{center}
\end{table}

\section{EXPECTED RESULTS}

The separation between pions and kaons in Cherenkov angle is given
by \cite{Jelly}
\begin{equation}
\label{eq:sep} \Delta\theta =
\cos^{-1}\left({1\over{n(\lambda)\beta_{\pi}}}\right)
-\cos^{-1}\left({1\over{n(\lambda)\beta_{K}}}\right)~~,
\end{equation}
where $n(\lambda)$ is the wavelength dependent index of refraction
and $\beta_i$ is the particles momentum divided by its energy.

 Thus determination of the expected separation requires
knowledge of the index of refraction, properly weighted by the
wavelength acceptance and the fact that Cherenkov light is generated
with a $1/\lambda^2$ spectrum.

We have measured the refractive indices of C$_4$F$_8$O,
C$_4$F$_{10}$ and also C$_4$F$_8$ at three different wavelengths
using a Michelson interferometer. The results are shown in
Fig.~\ref{index}. The curve is an analytical expression for the
C$_4$F$_{10}$ index based on extrapolating DELPHI measurements of
the C$_4$F$_{10}$ index in the ultraviolet region of the spectrum
\cite{indexC4F10}. We use this curve to model the wavelength
dependence of the index of refraction as we do not have measurements
over all the detector bandwidth, which extends down to 280 nm. We
are mostly insensitive to the actual value of the index, but need to
model the wavelength dependence to estimate the chromatic
aberration.

\begin{figure}[htb]
\centerline{\epsfxsize 4in
\epsffile{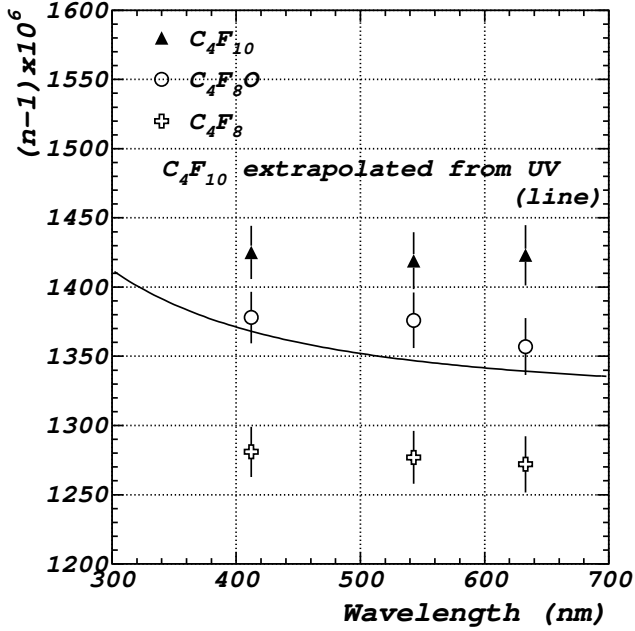}}
\vspace{-0.6cm} \caption{\label{index} Our measurement of the
index of refraction of three gases corrected to 1 atmosphere
pressure and 22$^{\circ}$C nominal temperature. The curve is an
extrapolation of C$_4$F$_{10}$ measurements at shorter wavelengths
that is used in the simulation.}
\end{figure}

The ability to identify particles depends not only on their inherent
separation in Cherenkov angle, but also upon the resolution provided
by the system. We model the expected resolution by using a full
Monte-Carlo simulation of the radiation and detection of Cherenkov
photons. Photons are generated according to the known Cherenkov
formula including polarization \cite{Jelly}. Wavelength dependence
of the refractive index is taken into account. The transmission and
reflection of all optical elements are taken into account using a
ray tracing method. This includes the wavelength dependent
transmission of the acrylic window. The quantum efficiency of the
MAPMTs as a function of wavelength is shown in Fig.~\ref{qef}; it
includes the cutoff due to the boroscilicate window. The curve is
based on Hamamatsu measurements. The collection efficiency of
electrons to the first dynode reduces the yield by an additional
77\%. The gas transmission is assumed to be 100\% in our wavelength
region. Using these numbers as input, for a 3 meter long C$_4$F$_8$O
radiator at room temperature, we expect to detect 46.3
photoelectrons for a high momentum track with an angular resolution
per photon of 0.75 mr. Therefore, after averaging over all the
photons, a resolution in Cherenkov angle per track of 0.103 mr is
expected, giving 4.2$\sigma$ separation between 70 GeV/c pions and
kaons. We note that the resolution has almost equal contributions of
0.5 mr from the index of refraction variation with wavelength
(chromatic abberation), the spatial error in detecting the photons
and the emission point error, which exists even in this focusing
system and is magnified by the fact that the mirror is tilted by 262
mr in order to focus on the MAPMT detector plane.

\begin{figure}[htb]
\vspace{-.5in} \centerline{\epsfxsize 3.0in \epsffile{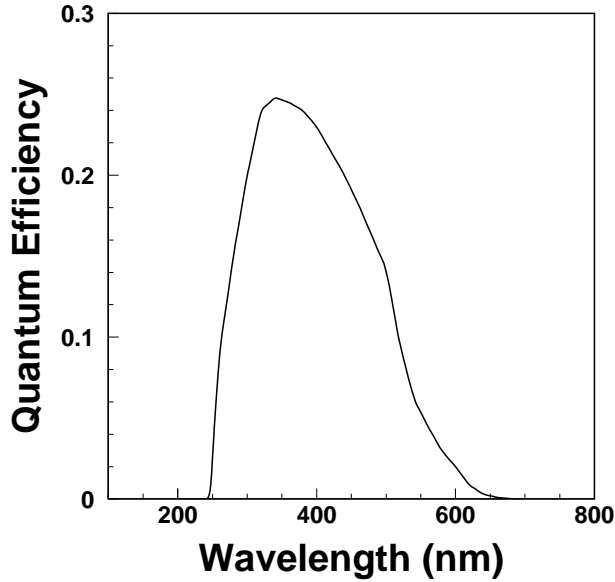}}
\vspace{-1.5cm} \caption{\label{qef} The quantum efficiency versus
wavelength of the MAPMT.}
\end{figure}

For our test situation the number of photons will be somewhat less
because of imperfect coverage of our photon detector and also
because the gas will not be totally purified. For our measured gas
mixture of (94$\pm$2)\% C$_4$F$_8$O and 6\% Argon, and a phototube
array coverage of 93.6\%, we expect 40.5 detected photons each with
a resolution of 0.75 mr, yielding an expected track resolution of
0.11 mr.

\section{OPERATING POINT DETERMINATION: HIGH VOLTAGE AND THRESHOLD OPTIMIZATION}

While in principle it should be able to operate the MAPMT system at
substantial gain, we find that large pulses can saturate the
electronics causing rather large voltage spikes to be reflected back
on the input cables between the MAPMT and the hybrids.  These sharp
voltage swings then capacitively couple to neighboring lines causing
extra hits in these channels, a phenomenon that we call
``cross-talk." Thus determining an optimal high voltage for the
system is very important. This problem is addressed in a subsequent
version of the VA\_MAPMT ASICs which have their dynamic ranges
optimized for higher gains. For the rare remaining residual
saturation occurrences, using a cable with less cross-talk would
also be beneficial.

Although each of our 53 tubes has a somewhat different gain as a
function of voltage, it was adequate to group them into three
categories: low gain, medium gain and high gain, designated as HV1,
HV2, HV3, respectively.

To investigate the performance as a function of high voltage we took
data at different settings. Ring images appeared at each setting.
Fig.~\ref{evt} shows a typical event. With our nominal threshold
setting little if no noise is evident.

\begin{figure}[htb]
\centerline{\epsfxsize 5.0in \epsffile{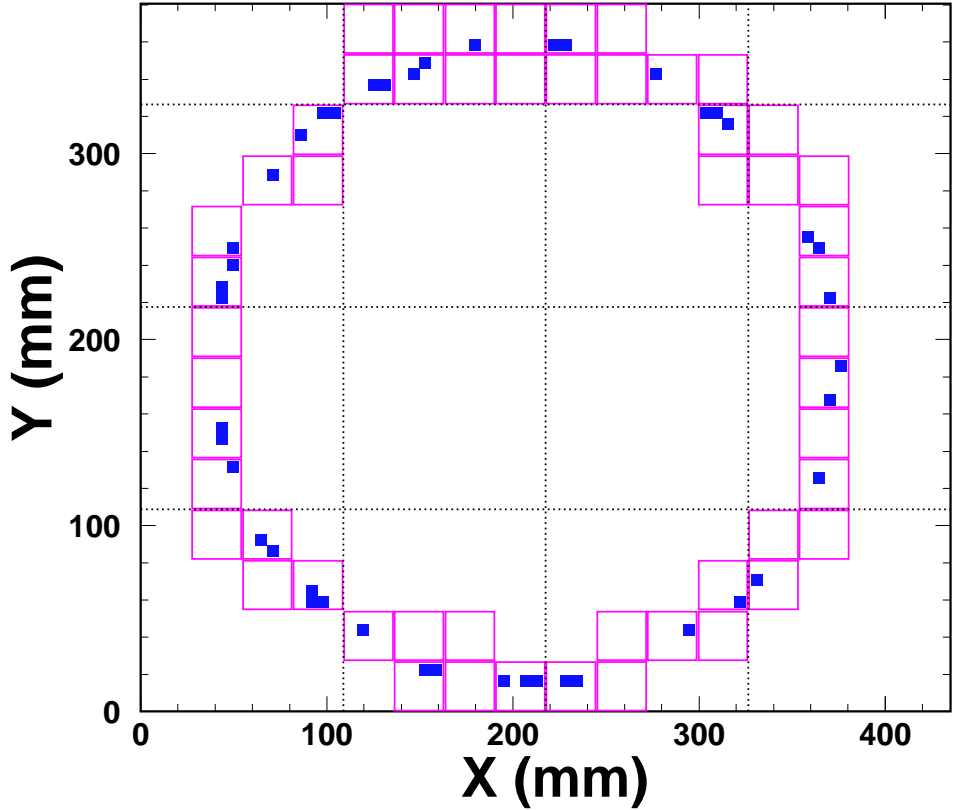}}
\vspace{-2.2mm} \caption{\label{evt} One recorded ring image. Each
darkened square represents a hit channel. Each solid hollowed box
corresponds to a MAPMT tube.}
\end{figure}

The photon yield can be characterized by counting the number of hit
channels. We also define a ``cluster" by grouping hits together if
they are in adjacent electronic channels. We are not grouping here
necessarily the channels on the MAPMT surface. What we are doing is
grouping the channels only by electronic address which is one
dimensional rather than two dimensional. While this does inevitably
group some real photons together it provides a useful way, in the
presence of cross-talk, of searching for a region where our
performance is independent of high voltage, usually called a
plateau. We also use the average cluster size.

Fig.~\ref{hv_scan} shows the number of hits, the number of clusters
and the average cluster size for each of the three groupings of
tubes as a function of high voltage. We choose to use 800 V, 750 V
and 700 V as nominal settings for the three groups, respectively.

\begin{figure}[htb]
\centerline{\epsfxsize 1.9in
\epsffile{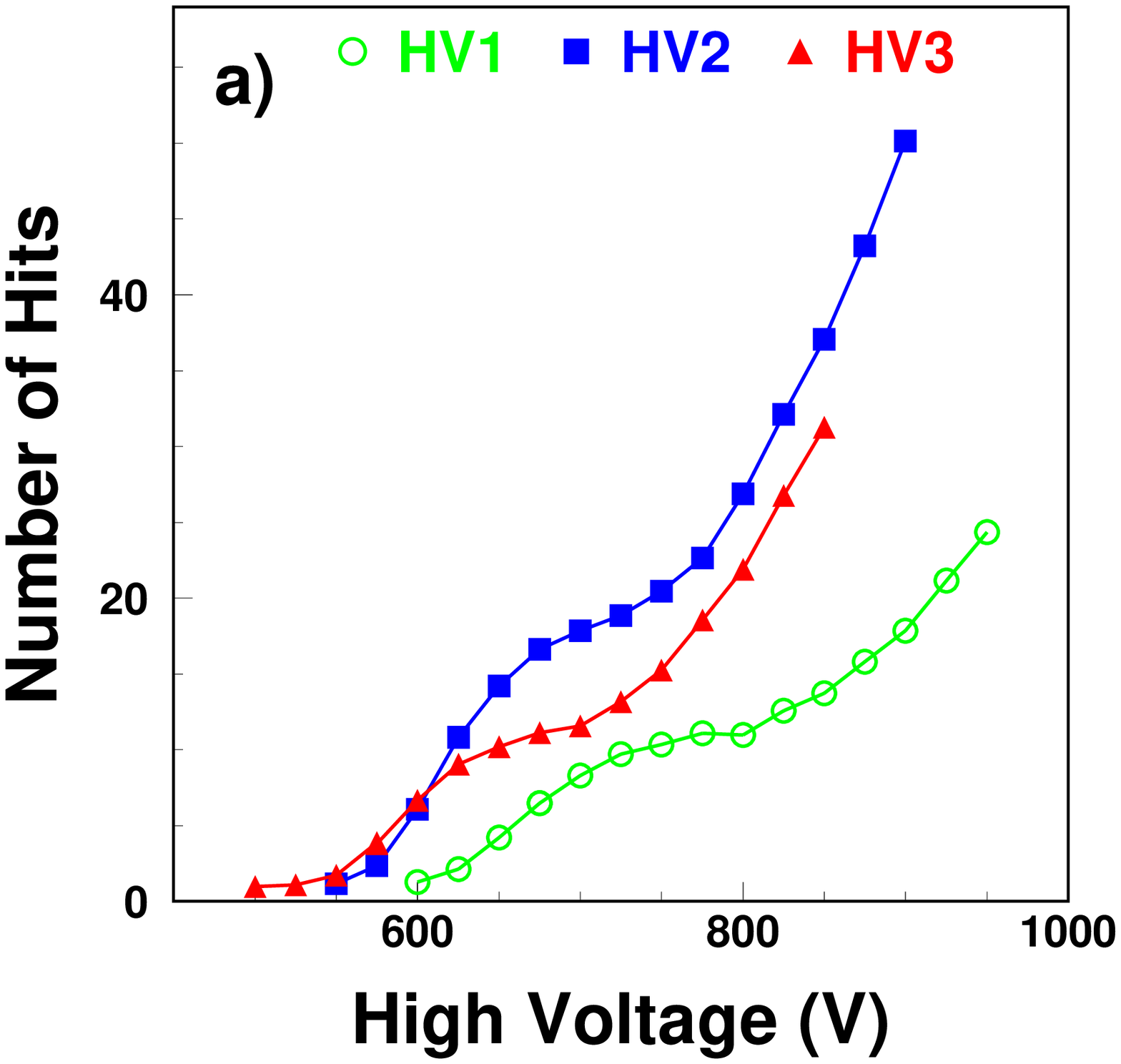}
            \hspace{0.1in}
            \epsfxsize 1.9in \epsffile{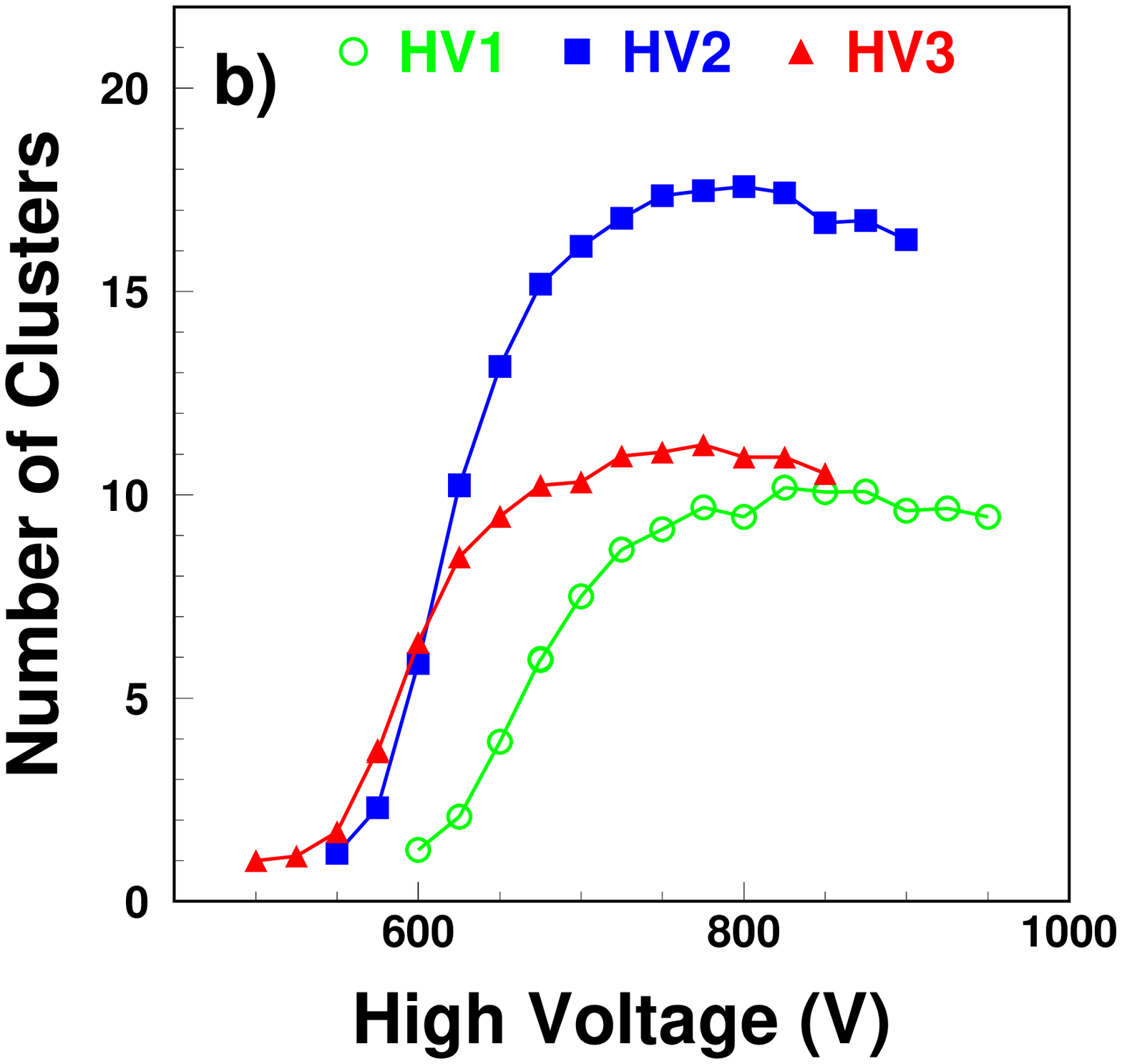}
            \hspace{0.1in}
            \epsfxsize 1.9in \epsffile{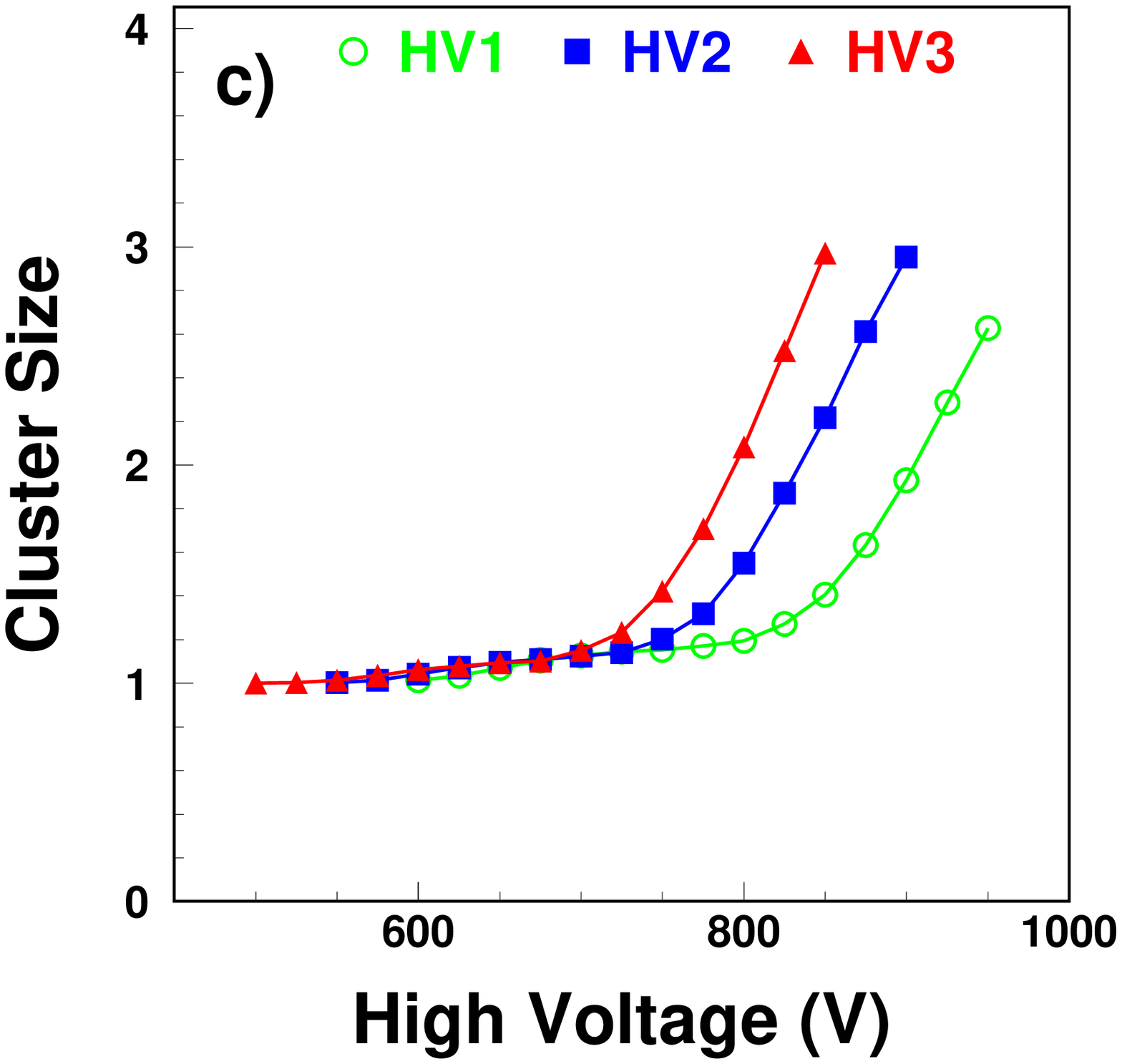}}
            \vspace{-0.3in}
\caption{\label{hv_scan} a) The number of hits, b) the number of
clusters and
   c) the average cluster size for each of the three groupings of MAPMT tubes
   as a function of high voltage.}
\end{figure}

A blue LED was employed to independently evaluate the size of the
cross-talk induced effects. The setting was such that only zero or
one photon was typically detected on any single tube in each pulse.
The cluster size as a function of voltage is shown in
Fig.~\ref{led_scan}. We estimate that at our nominal voltage setting
only 5\% of the observed photons are due to cross-talk.

\begin{figure}[htb]
\centerline{\epsfxsize 3.0in
\epsffile{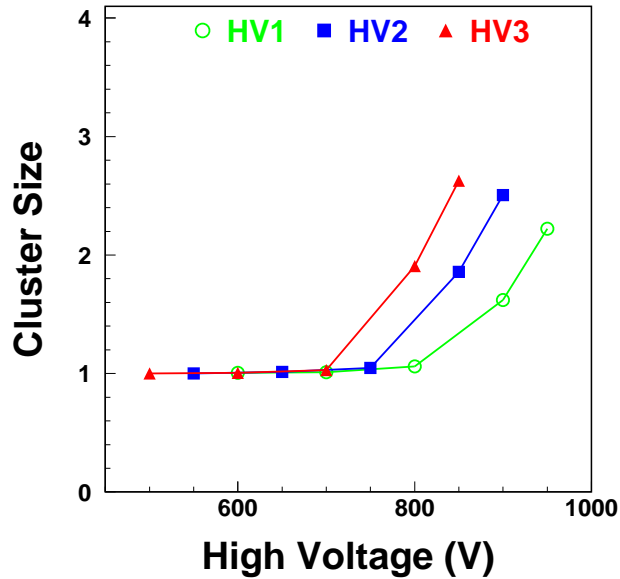}}
\vspace{-0.6in} \caption{\label{led_scan} The average cluster size
for each of the three groupings
   of MAPMT tubes as a function of high voltage, using the LED pulser instead of the proton beam. }
\end{figure}

The threshold setting controls the level for which signals are
counted. In Fig.~\ref{thr_scan} we show the variation in the number
of hits, number of clusters and cluster size as a function of
threshold from beam data. (The threshold voltage is negative; more
negative values result in higher thresholds.) The nominal threshold
is  -7.85 mV which corresponds to 49,100 electrons. This threshold
is at least ten times smaller than the mean charge expected from a
single photon. We tried to use a lower threshold, but did not see
any efficiency improvement. Our results are not very sensitive to
threshold which is kept as it was for the previously mentioned
studies (-7.85 mV) unless specifically indicated.

\begin{figure}[htb]
\centerline{\epsfxsize 1.9in
\epsffile{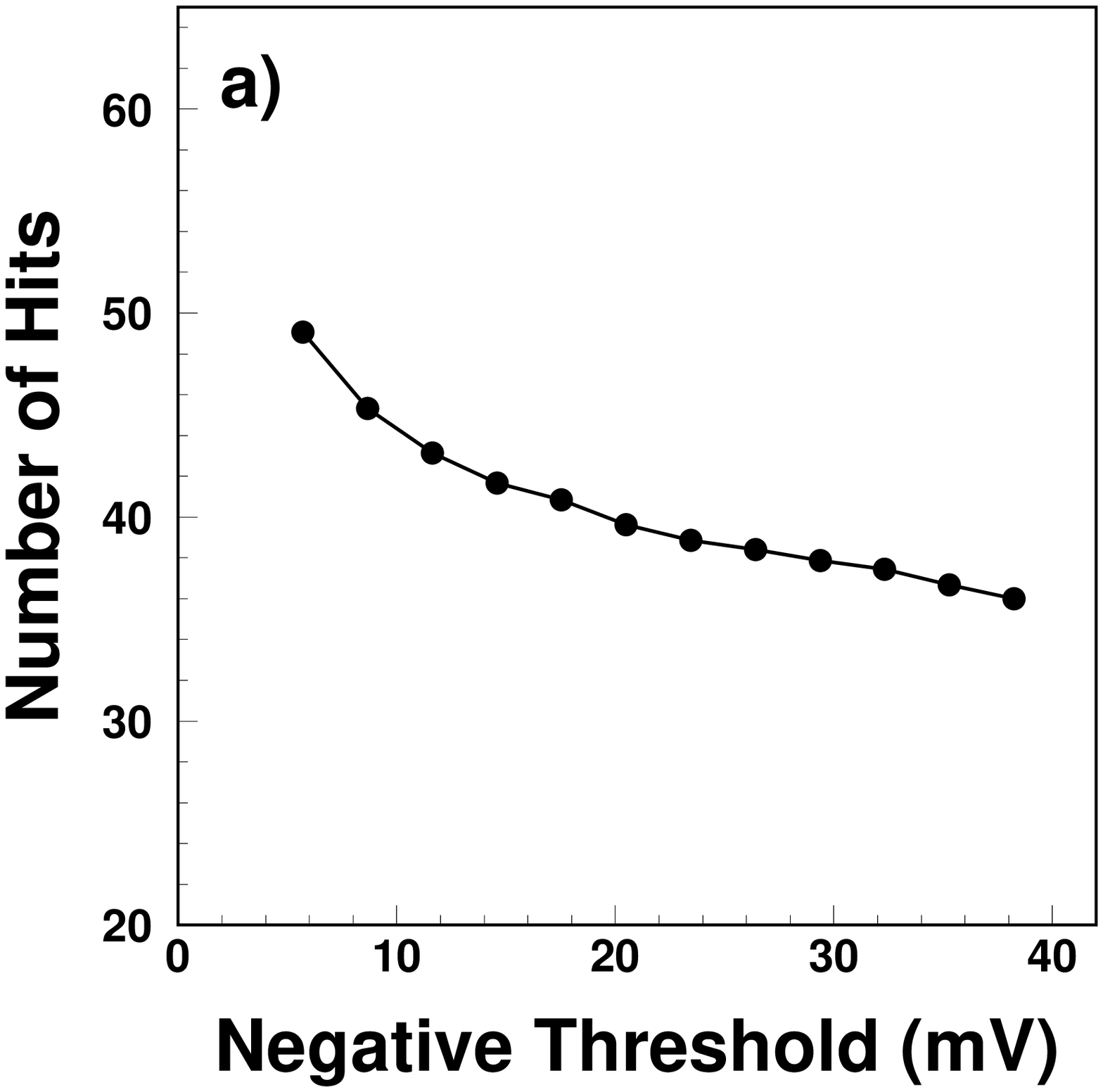}
            \hspace{0.1in}
            \epsfxsize 1.9in \epsffile{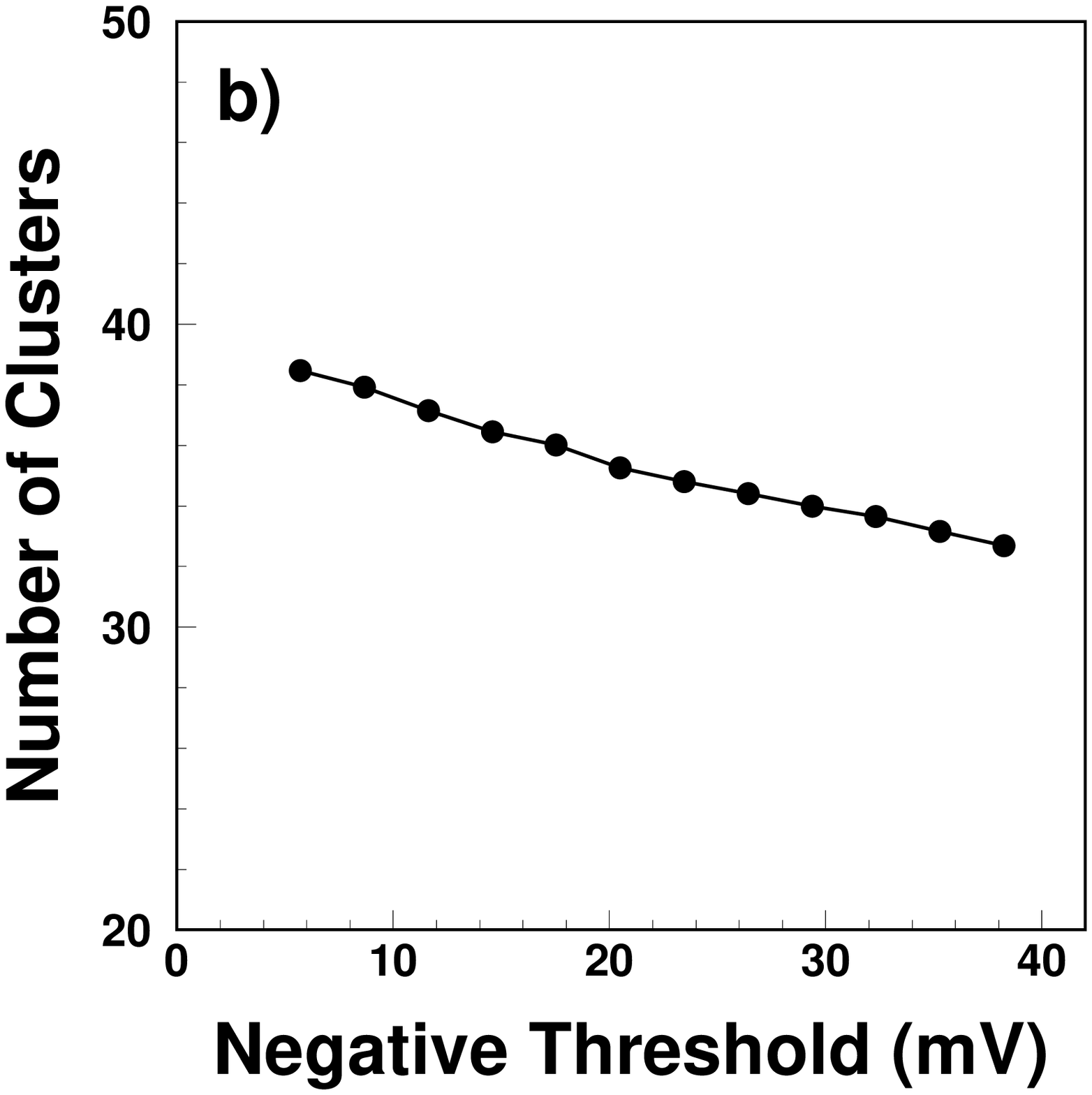}
            \hspace{0.1in}
            \epsfxsize 1.9in \epsffile{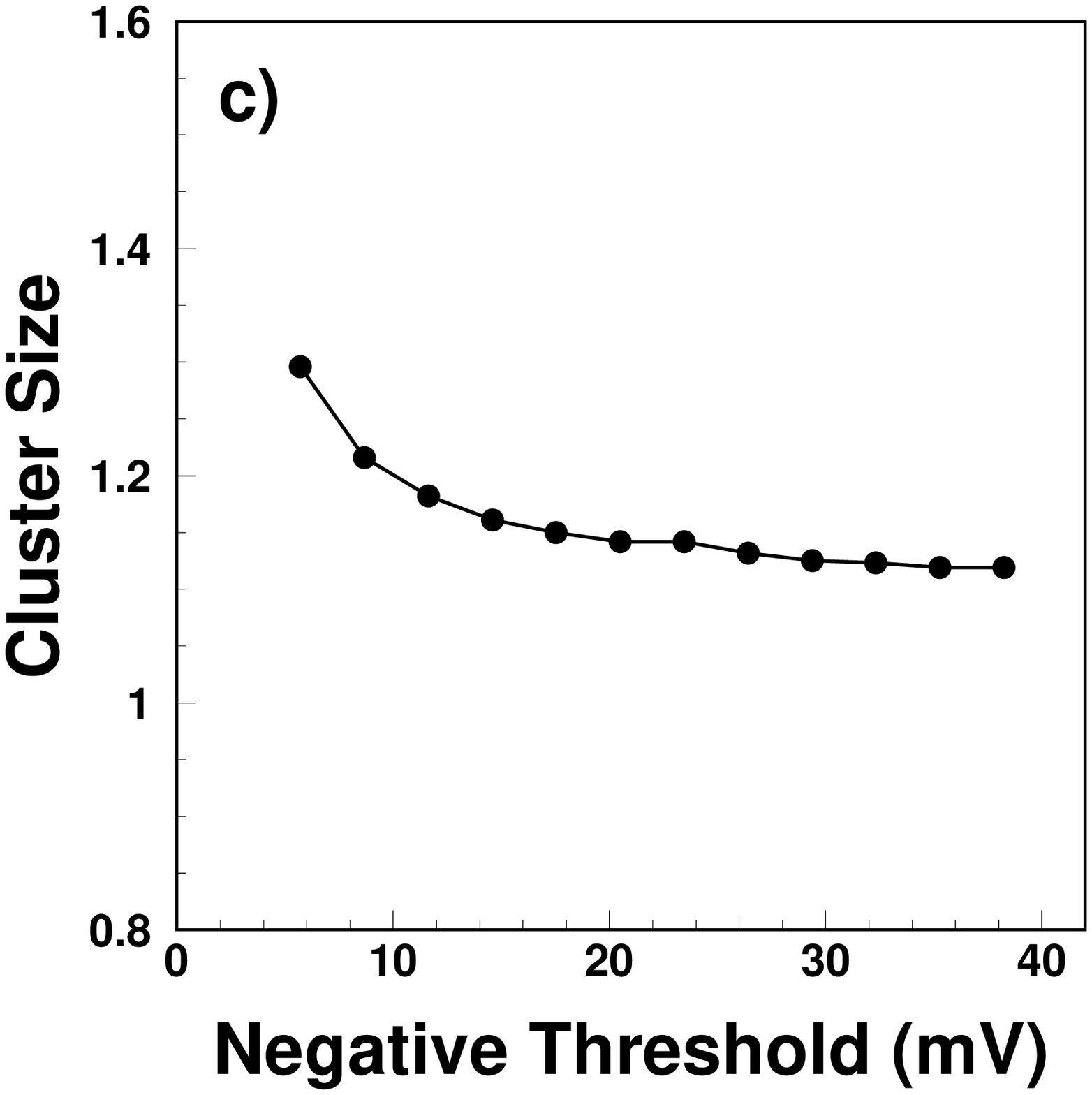}}
            \vspace{-0.3in}
\caption{\label{thr_scan} a) The number of hits, b) the number of
clusters and
   c) the average cluster size as a function of threshold in mV.
   The lowest threshold of 4.2 mV corresponds to about 26,200 electrons.}
\end{figure}

\section{RESULTS}

\subsection{Refractive Index}
We determine the effective refractive index from a fit to the
diameter of the Cherenkov rings, after correction to nominal
pressure (1 atm) and temperature ($22^{\circ}$C). After correcting
for the (6$\pm$2)\% of Argon, we find that the index of refraction
for C$_4$F$_8$O is 1.001389$\pm$0.000024 for the wavelength interval
of 280 - 600 nm weighted by the $1/\lambda^2$ input Cherenkov light
distribution, the window transparencies and the MAPMT efficiencies.
The error is dominated by the uncertainty in the gas purity. The
value obtained here is consistent with the measurements obtained
using the Michelson interferometer (see Fig.~\ref{index}).

\subsection{Photon Yield and Cherenkov Angular Resolutions}
We count photons about the mean Cherenkov angle for photons
generated by the track, each within three standard deviations of the
single photon resolution. The photon yield for an ensemble of tracks
is shown in Fig.~\ref{nhits_964}. The large shaded peak corresponds
to single tracks with a mean value of 43.1 photons, having a
Gaussian r.m.s. of 6.5 photons. The second peak is due to two tracks
arriving close in time. Although we predict 86.2 photons in this
peak, a Monte Carlo simulation taking into account that the tracks
are very closely correlated in both space and angle shows that we
expect only 76.9 photons, due to overlapping photons, consistent
with our observations. There is also a small tail toward low hit
numbers due to bad triggers or event fragments.

\begin{figure}[htb]
\centerline{\epsfxsize 3.0in
\epsffile{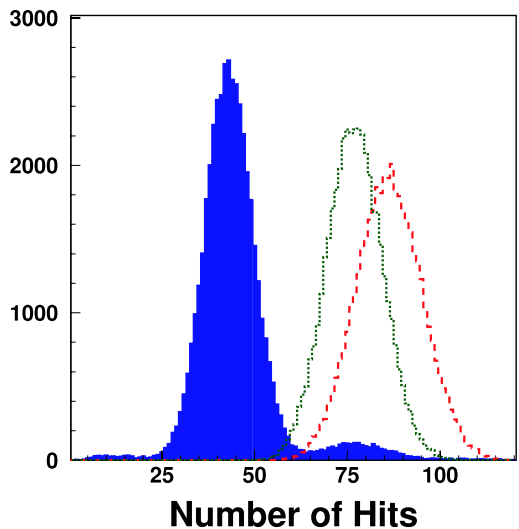}}
\vspace{-0.3in} \caption{\label{nhits_964} The photon yield per
track at nominal high voltage and threshold settings (solid). The
dashed curve shows the Monte Carlo expectation for two tracks
without considering overlap of photons, while the dotted curve
explicitly takes into account the overlap possibility.}
\end{figure}

The measured yield of 43.1 photons is larger than the Monte Carlo
expectation is 40.5 photons. The 6.5\% excess we observe is
consistent with the expected cross-talk, but may also be due to a
somewhat higher MAPMT quantum efficiency than expected.

The angular resolution per photon is shown in Fig.~\ref{res_photon}
(a). It is determined by plotting the measured Cherenkov angle for
each photon minus the expected Cherenkov angle, determined by
knowledge of the incident track angle. We measure an angular
resolution per photon of 0.79 mr while the Monte Carlo expectation,
shown in (b) is 0.75 mr. \cite{staterrs}. The difference is
consistent with being due to the $\sim$5\% cross-talk.

\begin{figure}[htb]
\centerline{\epsfxsize 2.5in
\epsffile{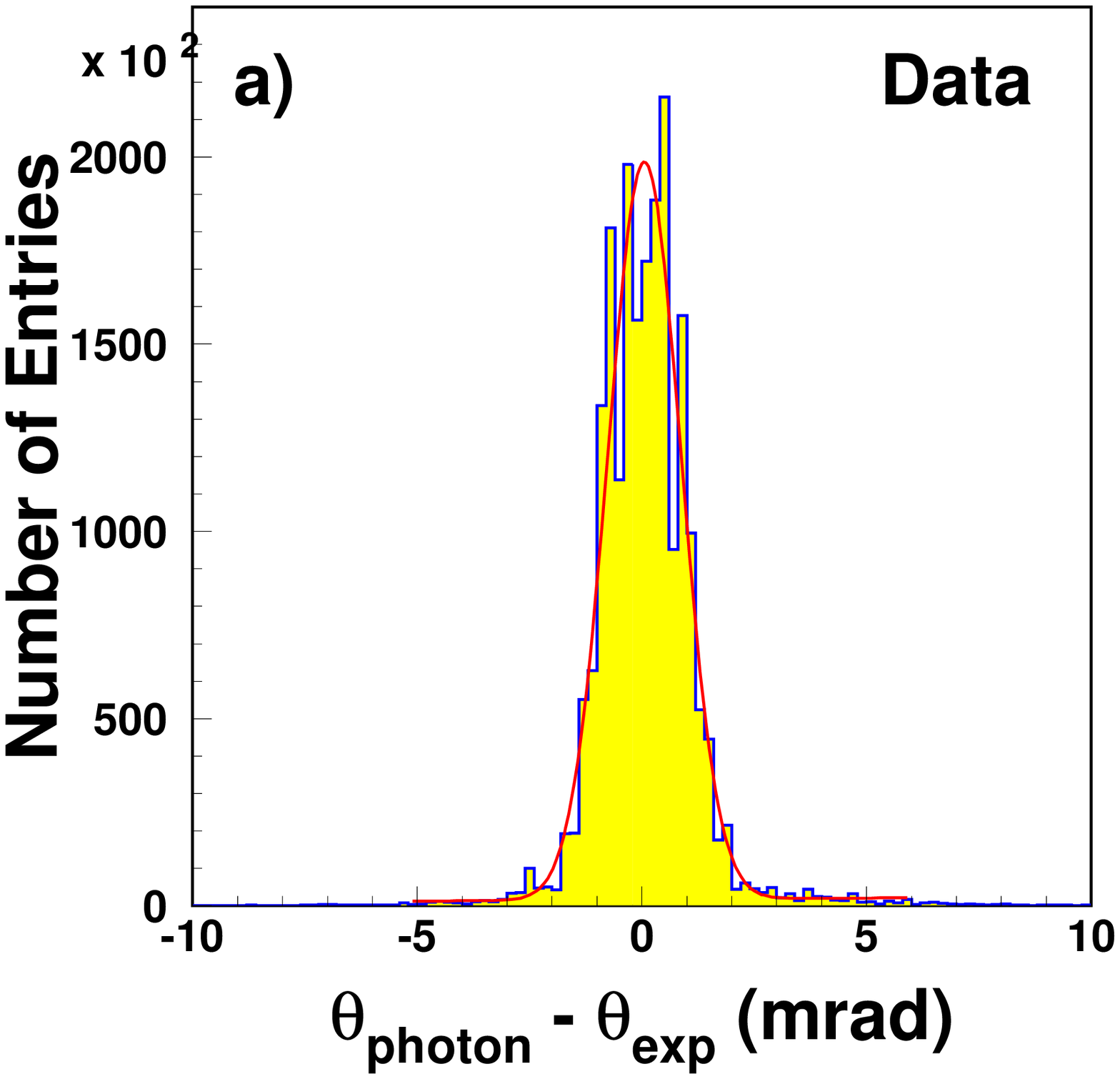}
            \hspace{0.1in}
            \epsfxsize 2.5in \epsffile{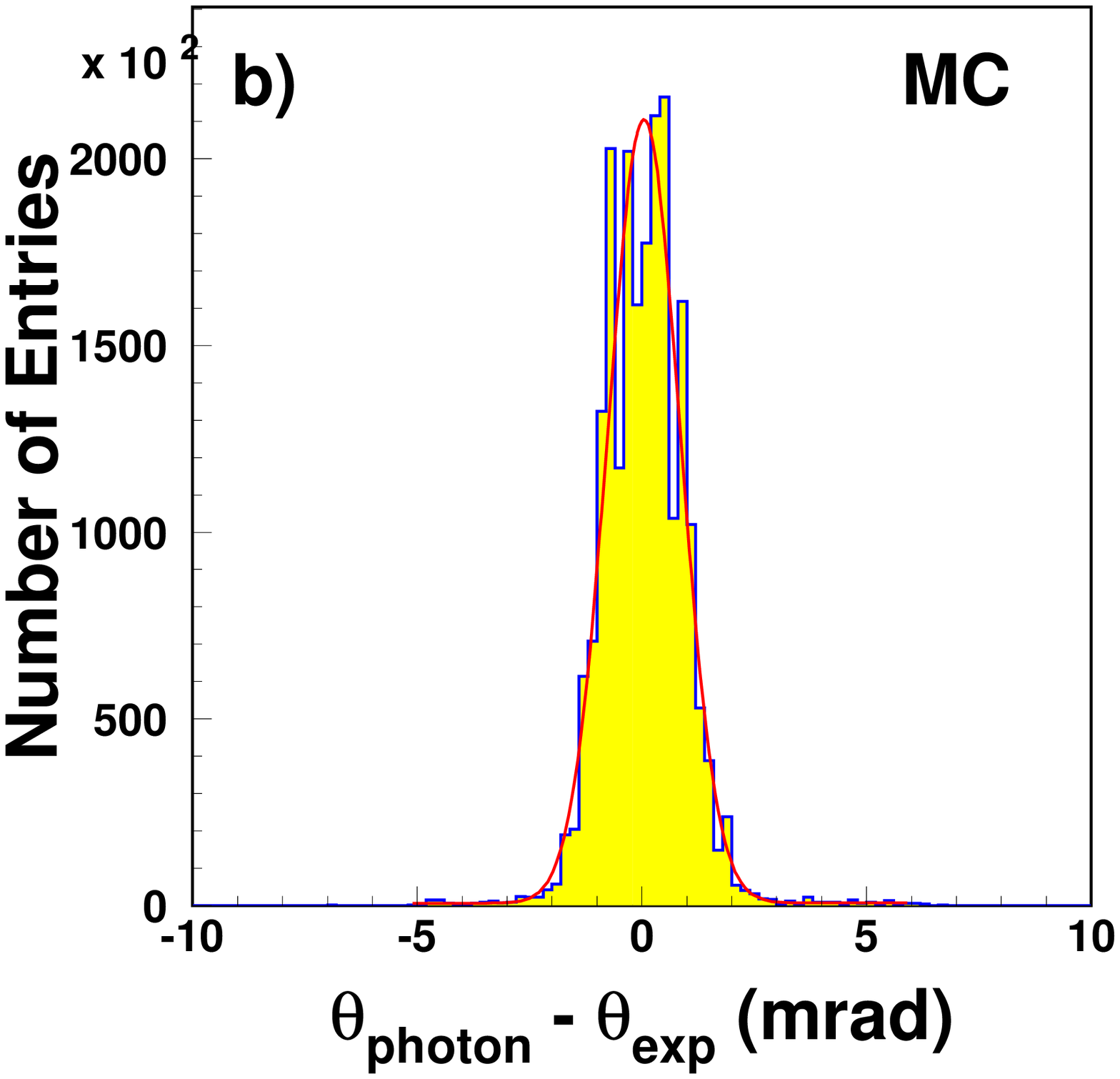}}
            \vspace{-0.4in}
\caption{\label{res_photon} The difference between measured and
expected Cherenkov angles for single photons in a) data and
   b) MC at nominal settings.}
\end{figure}

The Cherenkov angle per track is determined by averaging all
detected photons within three standard deviations of the single
photon resolution. In Fig.~\ref{run_964_rt_d} we show the measured
Cherenkov angle for each track minus the expected Cherenkov angle.
The angular resolution per track is 0.1164$\pm$0.0004 mr as
determined by a fit to a single Gaussian. It is slightly poorer than
the Monte Carlo expectation of 0.112 mr.

\begin{figure}[htb]
\centerline{\epsfxsize 3.0in
\epsffile{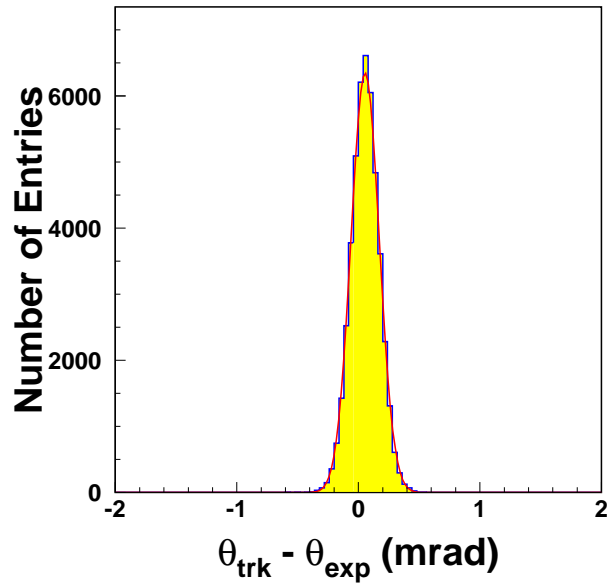}}
\vspace{-0.6in} \caption{\label{run_964_rt_d}The measured
Cherenkov angle minus the expected Cherenkov angle for each
track.}
\end{figure}

We now examine the track resolution as a function of high voltage.
Fig.~\ref{res_scan} shows the number of hits, the Cherenkov angular
resolution per photon and the Cherenkov angular resolution per track
as a function of HV2, the high voltage applied to group 2. All
groups are included here, however with HV1 being 50 V higher and HV3
50 V lower.

\begin{figure}[htb]
\centerline{\epsfxsize 1.9in
\epsffile{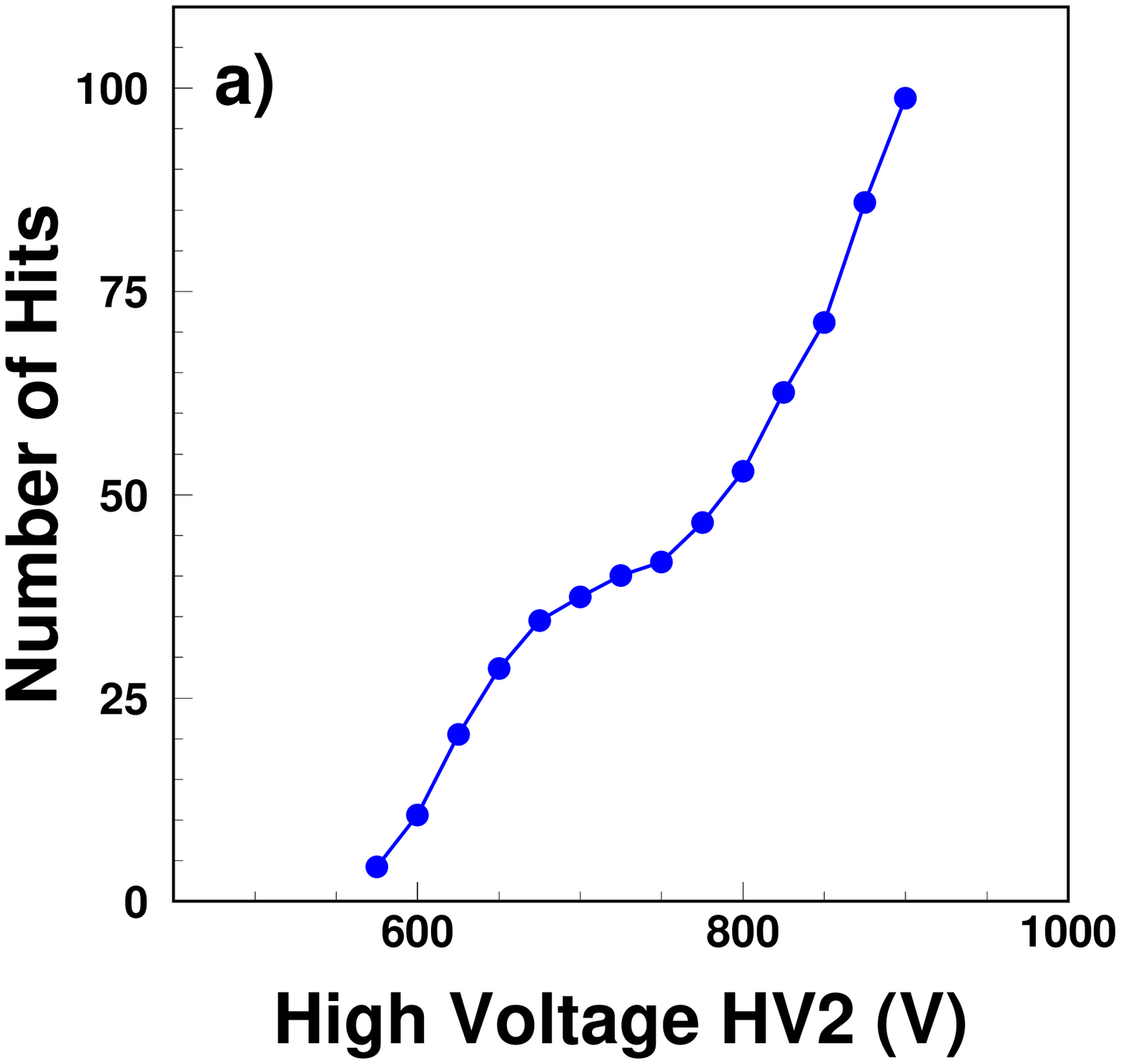}
            \hspace{0.1in}
            \epsfxsize 1.9in \epsffile{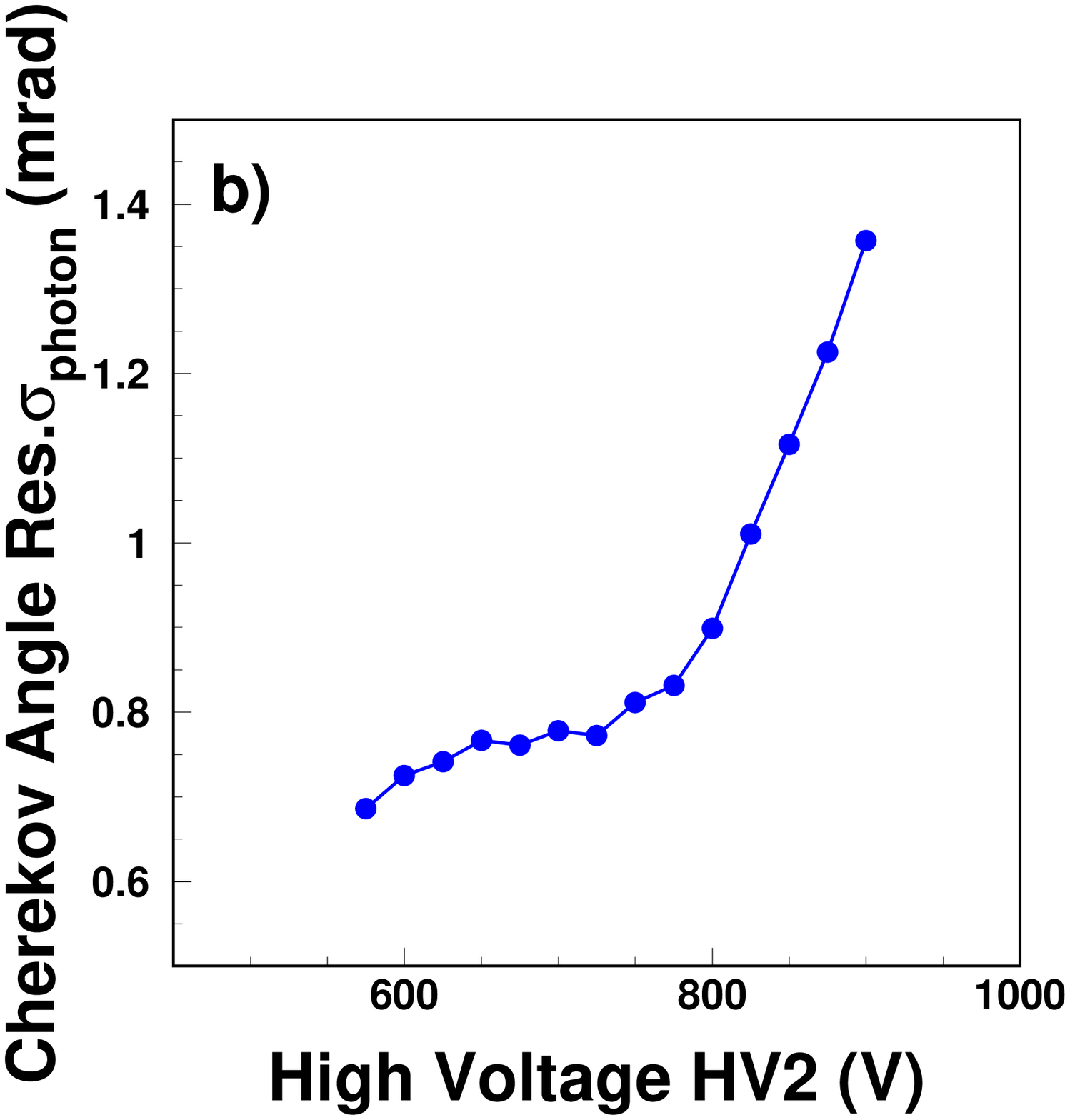}
            \hspace{0.1in}
            \epsfxsize 1.9in \epsffile{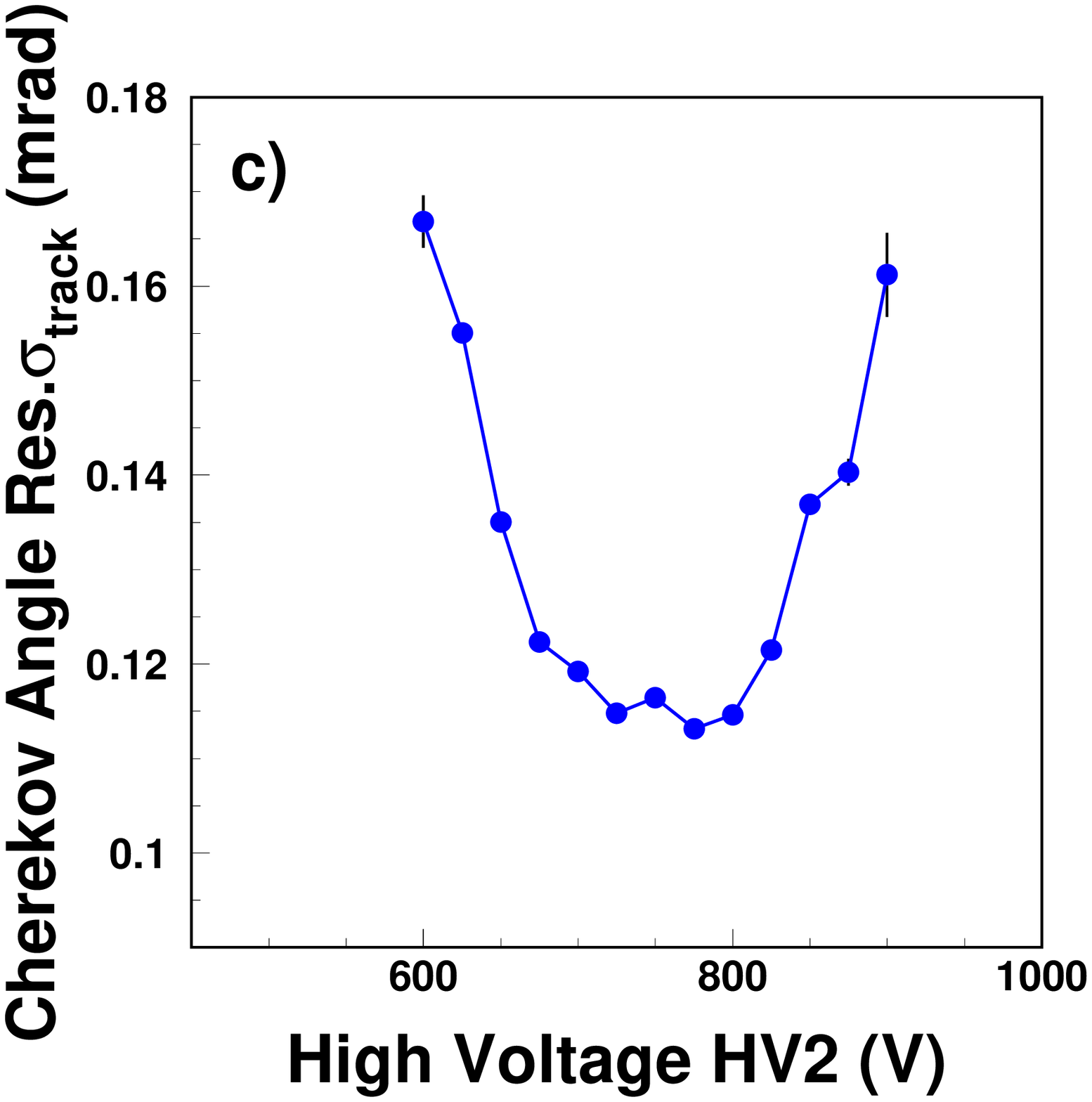}}
            \vspace{-0.3in}
\caption{\label{res_scan} a) The number of hits, b) per photon
Cherenkov angle resolution, and c) per track Cherenkov angle
resolution as a function of high voltage. Data are from all three
groups of MAPMT tubes. HV for the other two groups (HV1, HV3) are
set to be 50 V higher and 50 V lower respectively. }
\end{figure}

Viewed in terms of the angular resolution per track the voltage
plateau extends over a range of 100 V. Any higher or lower voltage
results in a poorer resolution. At the higher voltage this is due to
cross-talk and at the lower voltage a loss of hit efficiency. The
high voltage plateau can be extended by using the next generation of
front end ASICs and/or with better signal cables.

\section{CONCLUSIONS}

We have constructed a focussed ring imaging Cherenkov detector and
exposed it to a 120 GeV/c proton beam at Fermilab. Using a 3 meters
length of C$_4$F$_8$O gas radiator and R8900-M16 multianode
photomultiplier tubes we measure about 43.1 photons per track with
an angular resolution of 0.79 mr per photon and a track resolution
of 0.116 mr.  We expect that 5\% of the photons were generated by
cross-talk largely due to a mismatch between the dynamic range of
the front-end electronics and the gain of the MAPMTs. The gas
contained 6\% Argon and the photon detector covered 93.6\% of a
circle. In a final system the Argon would be eliminated and the
phototube coverage would be complete.

Based on these measurements, if we used pure gas and had complete
photon coverage, we would expect to have 46.6 photons and an angular
resolution per track of 0.109 mr. Our simulation using these
parameters gives the same number of photons and a slightly better
track resolution of 0.103 mr, since the simulation here doesn't
include the cross-talk. This system would be an excellent identifier
of particles.

Sometimes the parameter $N_0 = N_{\gamma}/L\sin^2\theta$, where
$N_{\gamma}$ is the number of detected photons, $L$ the detector
length and $\theta$ the Cherenkov angle is used to characterize RICH
systems \cite{Tomo}. This system would have an $N_0$ of 58/cm. The
usefulness of a RICH system, however, is given by its ability to
separate particles. $N_0$ is only part of the system design. It can
be increased, for example, by using photons at shorter wavelengths
but the chromatic abberations associated with such a change would
make the particle separation properties of the system worse, rather
than better. At 70 GeV/c incident particle momentum the angular
separation between pions and kaons in C$_4$F$_8$O is 0.421 mr. This
system would provide a separation corresponding to 4$\sigma$
separation.

    The R8900-M16 MAPMT from Hamamatsu has been shown to be an excellent
photon detector, especially when coupled to low noise electronics.
Bench tests confirm that the MAPMT has an active area of about 80\%
and the response over this area is quite uniform. For the four
central channels about 20\% of the photons are detected in one of
the adjacent channel. The edge and corner channels have
approximately a 15\%, and 10\% out-of-channel detection rate,
respectively. This results in an acceptably small degradation in the
positional resolution of the device. Lastly, we have found that
these tubes can be adequately shielded for operation in environments
which have magnetic fields. Shielding against the longitudinal
component requires extending the tube beyond the face of the MAPMT.
As long as photons are not incident at large angles, the loss due to
shadowing is generally small (it was found to be at the level of 5\%
for the BTeV RICH).

Lastly we found that \CFO\ is an excellent gas for Cherenkov
detectors with a relatively large refractive index, and appears to
be compatible with most materials.

 \section*{ACKNOWLEDGEMENTS}

We thank the U. S. National Science Foundation without whose support
this work would not have been possible. We also thank the Dr. E.
Ramberg and the Fermilab Beams Division for providing the test beam,
H. Cease of Fermilab for critical engineering support, L. Uplegger
for help with the data acquisition system and Terry Tope for help
with gas system programming. We thank Andre Braem for measuring the
mirror reflectivity.


\begin{thebibliography}{99}
\bibitem{thanks-brem}
The mirror reflectivity was measured in the center of the mirror. We
thank Andre Braem for making these measurements.
\bibitem{staterrs}
The statisical errors are very small $\pm$0.0004 mr on the measured
or simulated angles due to the large statistics.
\end{thebibliography}
\end{document}